\documentclass[%
 aip,
 jmp,%
 amsmath,amssymb,
 reprint,%
]{revtex4-1}

\usepackage{graphicx}
\usepackage{dcolumn}
\usepackage{bm}

\usepackage{color}
\begin{document}

\preprint{AIP/123-QED}

\title{High-Energy Radiation and Pair Production by Coulomb Processes in Particle-In-Cell Simulations}

\author{B. Martinez}\email{bertrand.martinez8@gmail.com}
\affiliation{CEA, DAM, DIF, F-91297 Arpajon, France}
\affiliation{CELIA, UMR 5107,Universit\'e de Bordeaux-CNRS-CEA, 33405 Talence, France}
\affiliation{LULI-CEA-CNRS, \'{E}cole Polytechnique, Institut Polytechnique de Paris, F-91128 Palaiseau, France}

\author{M. Lobet}
\affiliation{Maison de la Simulation, CEA, CNRS, Universit\'e Paris-Sud, UVSQ, 
Universit\'e Paris-Saclay, F-91191 Gif-sur-Yvette, France}

\author{R. Duclous}
\affiliation{CEA, DAM, DIF, F-91297 Arpajon, France}

\author{E. d'Humi\`{e}res}
\affiliation{CELIA, UMR 5107,Universit\'e de Bordeaux-CNRS-CEA, 33405 Talence, France}

\author{L. Gremillet}\email{laurent.gremillet@cea.fr}
\affiliation{CEA, DAM, DIF, F-91297 Arpajon, France}

 \email{email.com}

\date{\today}

\begin{abstract}

We present a Monte Carlo implementation of the Bremsstrahlung, Bethe-Heitler and Coulomb Trident processes into the particle-in-cell (PIC) simulation framework.
In order to address photon and electron-positron pair production in a wide range of physical conditions, we derive Bremsstrahlung and Bethe-Heitler cross sections taking
account of screening effects in arbitrarily ionized plasmas. Our calculations are based on a simple model for the atomic Coulomb potential that describes shielding due to
both bound electrons, free electrons and ions. 
We then describe a pairwise particle interaction algorithm suited to weighted PIC plasma simulations, for which we perform several validation tests. Finally, we carry
out a parametric study of photon and pair production during high-energy electron transport through micrometric solid foils. Compared to the zero-dimensional model of
J.~Myatt \emph{et al.} [Phys. Rev. E {\bf76}, 066409 (2009)], our integrated one-dimensional simulations pinpoint the importance of the electron energy losses resulting
from the plasma expansion.
\end{abstract}

\maketitle

Continuous progress in laser technology now makes available high-energy ($0.1-1$ $\mathrm{kJ}$), short duration ($0.1-1$ $\mathrm{ps}$) pulses, yielding
focused intensities in excess of $10^{20}\,\mathrm{Wcm}^{-2}$. Such laser parameters may give rise to a regime of laser-matter interaction where collective
plasma processes are coupled with strong radiation and electron-positron ($e^-e^+$) pair production \cite{RMPDiPiazza2012}. Most of the theoretical studies
conducted in past years have addressed the impact of the synchrotron photon emission and Breit-Wheeler pair production that result from the interaction
of the laser field with, respectively, high-energy electrons and photons, and are expected to prevail at ultra-high laser intensities ($I_L \gtrsim 10^{22}\,\rm Wcm^{-2}$)
\cite{PRLBell2008, PRLNerush2011, PRLRidgers2012, PRLBrady2012, PoPJi2014, PRLLobet2015, PoPGrismayer2016, PoPKostyukov2016,PRABLobet2017}. While such extreme conditions
should be achieved by the upcoming multi-petawatt laser systems (\emph{e.g.} CILEX-Apollon \cite{CLEOPapadopoulos2019}, CoReLS \cite{OLSung2017}, CAEP-PW\cite{OLZeng2017},
ELI \cite{MREWeber2017}), current experiments operate at significantly lower intensities, and are mostly prone to trigger Bremsstrahlung photon emission and Bethe-Heitler
(or Trident) pair production which, instead of the laser field, are mediated by the Coulomb field of atomic nuclei \cite{RMPKoch1959,PRSBethe1934,PRSBhabha1935}.

The incoherent Bremsstrahlung spectra originating from the interaction of laser-driven fast electrons with matter is a well-known feature of high-intensity
laser-solid experiments, which can serve for fast-electron characterization or radiography purposes \cite{PRLKmetec1992, PoPSchnurer1995, RSIPerry1999, PRLCowan2000, PRLSantala2000, PRLLedingham2000, PRLGlinec2005, NJPGaly2007, PoPChen2009, PoPWestover2010, PoPCompant2012, PoPCourtois2013}.
In the same context, the generation of $e^-e^+$ pairs directly follows from the Bremsstrahlung $\gamma$-ray photons interacting with heavy ions \cite{LPBCowan1999, APLGahn2000, PRLChen2009, *PRLCHen2010, *PRLChen2015, PoPWilliams2015, *PoPWilliams2016, HEDPWu2017}.
Record positron densities of $\sim 10^{16}\,\mathrm{cm}^{-3}$ have been reported using millimeter-sized high-$Z$ targets in which the Bethe-Heitler process
mainly accounts for pair production. These thick targets are either directly irradiated by intense picosecond lasers \cite{PRLChen2009} or penetrated by wakefield-driven
electron beams originating from a laser-irradiated gas jet \cite{PRLSarri2013, *NCSarri2015, PoPXu2016}. From measurements performed at various laser facilities, the
positron yield has been found to scale approximately quadratically with the laser energy, owing to increasingly energetic electrons when rising the laser intensity
and to their enhanced recirculation through the target \cite{PRLChen2015}. While quasineutral pair beams has been reported to be generated using wakefield-driven
relativistic electrons \cite{NCSarri2015}, the laser-solid experiments carried out so far have led to somewhat reduced $e^-e^+$ density ratios ($n_+/n_-\sim 0.5$),
as recently measured at the Texas Petawatt Laser Facility \cite{SRLiang2015}. 

The particle-in-cell (PIC) technique \cite{BOOKBirdsall2004} is widely used to simulate the kinetic and collective phenomena at play in intense laser-plasma interactions.
In anticipation of the future multi-PW laser experiments, much effort has been lately made worldwide to enrich PIC codes with numerical models describing synchrotron photon
emission and Breit-Wheeler pair production \cite{PPCFDuclous2011, JPCSLobet2016,PRLYi2016, PoPPandit2012, PoPGrismayer2016, PREGonoskov2015}. Since these processes are mediated
by electromagnetic fields, they mainly take place within the laser-irradiated region, and thus do not require increasing the typical space-time scales of the laser-plasma
simulation. By contrast, the Coulomb-field-mediated processes of radiation (Bremsstrahlung) and pair production (Bethe-Heitler, Trident) arise during the relaxation of
the laser-driven relativistic electrons through the dense target, and therefore over time ($\sim 1-10\,\rm ps$) and ($\sim \rm mm$) scales usually much larger 
than those characterizing the laser interaction. As this puts strong computational constraints on integrated PIC simulations, the radiation and pair production physics
in current laser experiments is typically modeled using dedicated Monte Carlo codes with input fast electron sources estimated from theoretical arguments, PIC simulations or
experimental data \cite{PRLChen2009, PRLSarri2013,ASSHenderson2011,PoPYan2012, *PoPYan2013a, EPJDJiang2014}.

Although a full-scale, self-consistent modeling of the Coulomb-field-mediated radiation and pair production in intense laser-solid interactions is still outside the reach of
multidimensional PIC codes, it remains worthwhile to study these effects over the restricted space-time scales currently accessible to simulations. This has motivated a
number of groups to implement Bremsstrahlung \cite{PoPSentoku1998, QEAndreev2010, CLFWard2014, PPCFVyskocil2018,HPLSEWu2018} and Bethe-Heitler
\cite{JPCSMoritaka2013, LPBNakamura2015} packages into PIC codes. All of these works employ a Monte Carlo (MC) approach based on analytical \cite{RMPKoch1959}
or tabulated \cite{ADNDTSeltzer1986} cross sections, which are expected to be most valid for isolated neutral atoms. Since intense laser-solid interactions may lead
to a variety of ionization states, and hence atomic screening effects, it is useful to provide more general cross sections in view of their implementation into PIC-MC laser-plasma
simulation codes. This is the first objective of the present article. The second one is to present a Monte Carlo pairwise interaction scheme, which is adapted to macro-particles
with arbitrary numerical weights. Finally, we apply our numerical model to the investigation of pair production during fast-electron transport within a self-consistent simulation framework.

This paper is organized as follows. In Sec.~\ref{sec:derivation}, by combining the theoretical formulas reviewed in Refs.~\onlinecite{RMPKoch1959, RMPMotz1969} with a mixed
Thomas-Fermi-Debye screened atomic potential \cite{PRANardi1978, *LPBNardi2007, JQSRTRozsnyai1979}, we derive a set of modified Bremsstrahlung and Bethe-Heitler cross sections
valid for partially ionized dense plasmas. These expressions, supplemented with the Coulomb Trident cross section \cite{PRSBhabha1935} for direct pair production, form the
basis of a Monte Carlo package included in the PIC-MC \textsc{calder} code, describing Coulomb-mediated radiation and pair production processes. Our numerical implementation,
detailed in Sec.~\ref{subsec:binary_collision}, relies on a macro-particle-pairing algorithm that handles arbitrarily weighted macro-particles, originally developed for modeling elastic Coulomb
collisions \cite{JCPNanbu1998, PoPPerez2012}. In Sec.~\ref{subsec:test} we demonstrate the capability of \textsc{calder} to accurately model the total (collisional-radiative)
electronic stopping power of a solid target in a broad range of electron energies, while in Sec.~\ref{sec:generation} we study the positron generation accompanying the relaxation
of fast electrons inside a solid copper foil of variable thickness. Finally, we summarize our results in Sec. ~\ref{sec:summary}.

\section{Bremsstrahlung and pair production cross sections in partially ionized plasmas} \label{sec:derivation}

In this first section, we obtain analytic cross sections for the electron Bremsstrahlung and Bethe-Heitler processes, taking account of Thomas-Fermi and Debye-type
screening effects in a unified fashion depending on the plasma parameters. 

\subsection{Simple atomic potential model} \label{subsec:screen_effect}

The Coulomb interaction between a high-energy electron and an ion's nucleus is modified by the screening due to bound electrons, free electrons and plasma ions, depending
on the ionization state of the medium. For neutral atoms of atomic number $Z$, the Coulomb potential around the nuclear charge can be assumed of the Yukawa type:
\begin{align}
V_\mathrm{TF}(r) &= \frac{q}{r}\exp\left(-r/L_{\mathrm{TF}}\right)\,, \\
L_\mathrm{TF} &=4\pi \epsilon_{0}\frac{\hbar^{2}}{m_{e}e^{2}} Z^{-1/3}\,, \label{eq:LTF}
\end{align}
where $q=Ze/4\pi\epsilon_0$ and $L_\mathrm{TF}$ is the Thomas-Fermi length accounting for shielding by bound electrons. We have introduced $\epsilon_0$ the permittivity
of free space, $m_e$ the electron mass, $e$ the elementary charge, and $\hbar = h/2\pi$ the Planck constant. More precise multi-exponential fits of the Thomas-Fermi
potential could be used \cite{ZNAMoliere1947}, but we will limit ourselves to the above simple approximation. While Eq.~\eqref{eq:LTF} applies, in principle, to an
isolated neutral atom (where charge neutrality is fulfilled at infinity), we assume that it also holds in a cold neutral medium (where charge neutrality is fulfilled
at the ion-sphere radius) \cite{APMarch1957}. 

In a highly ionized plasma, the Coulomb potential can be modeled in a similar form:
\begin{align}
V_\mathrm{D}(r) &= \frac{q}{r}\exp\left(-r/L_{\mathrm{D}}\right)\,, \\
L_\mathrm{D} &=\sqrt{\frac{\epsilon_0 k_\mathrm{B} T}{e^2 n_i Z^{*} \left(Z^{*}+1\right)}}\,, \label{eq:LDH}
\end{align}
where $L_\mathrm{D}$ is the Debye length that describes screening by free electrons and plasma ions, $Z^*$ is the ionization degree, $k_\mathrm{B}$ is the
Boltzmann constant and $n_i$ is the ion density. We have supposed a globally neutral plasma ($n_e=Z^*n_i)$ and equal electron and ion temperatures ($T_e=T_i=T$).
To address coupled plasma regimes, we impose a lower bound on $L_\mathrm{D}$, equal to the interatomic distance  \cite{POFLee1984} $r_i = (3/4\pi n_i)^{1/3}$.
In practice, the ionization degree $Z^*$ is evaluated using a numerical fit to the Thomas-Fermi model for a finite-radius atom \cite{AAMPMore1985}.

In the general case of a partially ionized plasma, following Refs.~\onlinecite{PRANardi1978,JQSRTRozsnyai1979}, we assume for simplicity that the Coulomb
potential can be described as a weighted sum of the above Thomas-Fermi and Debye screened potentials:
\begin{equation}
V_{\mathrm{TFD}}\left(r\right) = \frac{q_{\rm TF}}{r}\exp(-r/L_{\rm TF}) + \frac{q_{\rm D}}{r}\exp(-r/L_{\rm D}) \,,
\label{eq:V_TFD}
\end{equation}
with $q_{\rm TF}/q = 1-Z^{*}/Z$ and $q_{\rm D}/q = (Z^{*}/Z)$. In a cold neutral medium ($Z^*\rightarrow 0$), we have $V_\mathrm{TFD}\rightarrow V_\mathrm{TF}$,
whereas, in a fully ionized plasma, $V_\mathrm{TFD}\rightarrow V_\mathrm{D}$, as expected. Also, $V_\mathrm{TFD}(r)\rightarrow Ze/4\pi\epsilon_{0}r$ when $r \rightarrow 0$
as it should. More accurate screening models could be used \cite{PRADas2016} but at the expense of analytical simplicity.

\subsection{Bremsstrahlung cross sections} \label{subsec:cross_section_br}

\begin{figure}[tbh]
\centering
\includegraphics[width=0.48\textwidth]{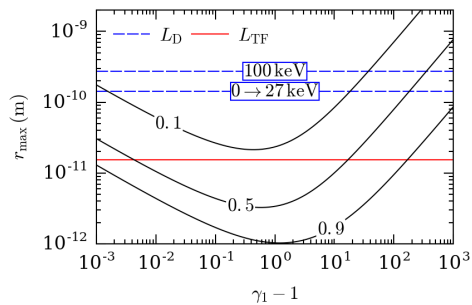}
\caption{Maximum impact parameter $r_\mathrm{max}$ (solid black curves) vs normalized electron kinetic energy for relative photon energies $k/(\gamma_1-1)\in (0.1,0.5,0.9)$.
The Debye screening length ($L_{\rm D}$) is plotted for a solid-density Cu plasma of temperature $T\in (0, 27)\,\rm keV$ and $T=100\,\rm keV$ (dashed blue lines). In the former
case, $L_{\rm D}$ is bounded by the interatomic distance. The Thomas-Fermi screening length ($L_{\rm TF}$) is plotted as a red line.
}
\label{fig:screening_study}
\end{figure}

Let us consider an electron of total energy $E_1/m_e c^2=\gamma_1$ and normalized momentum $p_1=\sqrt{\gamma_1^2-1}$, both measured in the ion rest frame. After emitting a
photon of normalized energy $k=\hbar \omega/m_{e}c^{2}$ in the screened atomic field, the normalized electron energy and momentum become $\gamma_2=\gamma_1-k$ and
$p_2=\sqrt{\gamma_2^2-1}$, respectively. Let us introduce $r_{\rm C}=\hbar /m_ec$ the Compton radius. For given electron and ion parameters, the importance of screening effects on Bremsstrahlung can be assessed by
comparing the maximum impact parameter, $r_\mathrm{max} = r_{\rm C} /(p_1-p_2-k)$ with the Thomas-Fermi and Debye screening lengths \cite{RMPKoch1959}. If $r_\mathrm{max}$
is much smaller than $L_\mathrm{TF}$ or $L_\mathrm{D}$, then the corresponding screening process is expected to be negligible. Figure~\ref{fig:screening_study} plots $r_\mathrm{max}$ as
a function of the normalized electron kinetic energy, $\gamma_1-1$, for various relative photon energies $k/(\gamma_1-1)\in (0.1,0.5,0.9)$. Overlaid are plots of $L_\mathrm{TF}$
and $L_\mathrm{D}$ in a solid-density copper plasma of variable temperature ($T\in (0-27)\,\mathrm{keV}$ and $T=100\,\mathrm{keV}$). At solid density, the Debye length
defined by Eq.~\eqref{eq:LDH} exceeds the interatomic distance ($r_{i}=1.4\times 10^{-10}\, \rm m$) only at temperatures $>27\,\rm keV$. From this graph, both Thomas-Fermi and
Debye shielding effects should be more pronounced at very high, and to a lesser degree, low electron energies; besides, at fixed electron energy, they should be enhanced with
decreasing photon energy.

In the following, photon-energy-differential Bremsstrahlung cross sections are derived based on the atomic potential \eqref{eq:V_TFD}. As to our knowledge
there is no general analytic theory for electron energies varying from the keV to the GeV ranges, we draw upon the results of Ref.~\onlinecite{RMPKoch1959},
and make use of three distinct formulae, respectively valid for (i) non-relativistic ($1< \gamma_1\lesssim 2$); (ii) moderately relativistic ($2\lesssim \gamma_1 \lesssim 100$) and (iii) ultra-relativistic
($\gamma_1\gtrsim 100$) electron energies.

\subsubsection{Non-relativistic electrons} \label{subsubsec:non_relativistic}

Introducing $\widetilde{V}_\mathrm{TFD}\left(\mathbf{u}\right)$ the Fourier transform of $V_{\rm TFD}(\mathbf{r})$, normalized by the factor $e/(\alpha_{f}m_{e}c^{2})$
\begin{equation}\label{eq:TF_LDH} 
 \widetilde{V}_\mathrm{TFD}(\mathbf{u}) =  \frac{1}{\left(2\pi\right)^3}\int \frac{eV_\mathrm{TFD}(\mathbf{r})}{\alpha_{f}m_{e}c^{2}} \exp \left(i \frac{\mathbf{u}\cdot\mathbf{r}}{r_{\rm C}}\right) \, d^{3}\left(\frac{\mathbf{r}}{r_{\rm C}}\right) \, ,
\end{equation}
the non-relativistic ($1<\gamma_1 \le 2$) electron Bremsstrahlung cross section, differential in the photon energy, writes in the Born approximation \cite{OUPHeitler1954}:
\begin{align}
\frac{d\sigma_\mathrm{Br,nr,TFD}}{dk} & = \frac{64 \pi^4 r_e^2\alpha_f}{3 k p_1^2} \int_{\delta p_-}^{\delta p_+}
\left \vert\widetilde{V}_\mathrm{TFD}(u)\right \vert^2 u^3 \, du \label{eq:csnr} \,.
\end{align}
Here $\alpha_f=e^2/(4\pi\epsilon_0 \hbar c)$ denotes the fine structure constant, $r_e=e^2/(4\pi\epsilon_0 m_e c^2)$ is the classical electron radius, and
$\delta p_+ = 2p_1-k$ and $\delta p_- = k$ are, respectively, the maximum and minimum momentum transfers to the nucleus in the collision.

After substituting Eqs.~\eqref{eq:V_TFD} and \eqref{eq:TF_LDH} into Eq.~\eqref{eq:csnr}, one obtains the non-relativistic Bremsstrahlung cross section
(see Appendix \ref{sec:app_cs_br_nr_deriv}):
\begin{align} \label{eq:cs_br_nr}
\frac{d\sigma_\mathrm{Br,nr,TFD}}{dk} &=  \frac{16r_e^2 Z^ 2\alpha_f}{3kp_1^2}\left[g\left(\delta p_{+},\delta p_{-},\eta_\mathrm{TF},\bar{L}_\mathrm{TF}\right)\right. \nonumber \\
&+\left.g\left(\delta p_{+},\delta p_{-},\eta_\mathrm{D},\bar{L}_\mathrm{D}\right)+\Gamma_{c}\right] \,,
\end{align}
where we have introduced the function $g$,
\begin{align}
g\left(\delta p_+,\delta p_-,\eta,L\right) &= \frac{\eta^2}{2} \left[ \ln \left( \frac{\delta p_+^2 \bar{L}^2+1}{\delta p_-^2 \bar{L}^2+1}\right) \right. \nonumber \\
& +\left.\frac{1}{\delta p_+^2 \bar{L}^2+1}-\frac{1}{\delta p_-^2 \bar{L}^2 +1} \right] \,,
\end{align}
and the coupling term $\Gamma_c$,
\begin{align}
\Gamma_c &= \frac{\eta_\mathrm{TF} \eta_\mathrm{D}}{(\bar{L}_\mathrm{D}^2 - \bar{L}_\mathrm{TF}^2)}
\left[ \bar{L}_\mathrm{TF}^2 \ln \left( \frac{\delta p_{-}^2 \bar{L}_\mathrm{D}^{2}+1}{\delta p_{+}^2 \bar{L}_\mathrm{D}^2 + 1}\right) \right. \nonumber \\
 & +\left. \bar{L}_\mathrm{D}^2 \ln \left( \frac{\delta p_{+}^2 \bar{L}_\mathrm{TF}^2 + 1}{\delta p_{-}^2 \bar{L}_\mathrm{TF}^2 + 1} \right) \right] \,.
\end{align}
with $\eta_\mathrm{TF}=q_{\rm TF}/q=1-Z^{*}/Z$, $\eta_\mathrm{D}=q_{\rm D}/q=Z^{*}/Z$ and $\bar{L}$ denotes lengths, normalized by the Compton radius: $\bar{L}=L/r_{\rm C}$.

\begin{figure}[tbh!]
\includegraphics[width=0.48\textwidth]{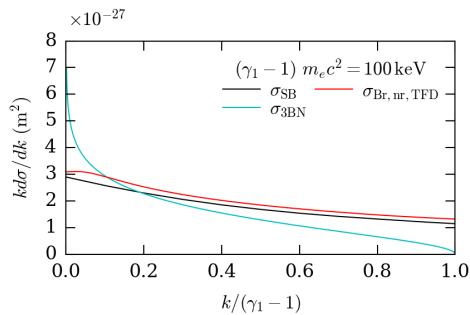}
\caption{Non-relativistic Bremsstrahlung cross section $kd\sigma/dk$ vs normalized photon energy $k/(\gamma_1-1)$ for a $100\,\mathrm{keV}$ energy electron interacting
with neutral Cu atoms. Comparison between the Elwert-corrected formula ($kd\sigma_\mathrm{Br,nr,TFD}/dk$, red curve), Seltzer and Berger's data \cite{ADNDTSeltzer1986}
($k d\sigma_\mathrm{SB}/dk$, black curve) and the nonscreened formula 3BN of Ref.~\onlinecite{RMPKoch1959} ($k d\sigma_\mathrm{3BN}/dk$, cyan curve).
\label{fig:cs_br_nr}}
\end{figure}

As is well-known \cite{ADPElwert1939}, the accuracy of the Born approximation can be improved in the non-relativistic regime by multiplying Eq.~\eqref{eq:cs_br_nr}
by the Elwert correction factor \cite{ADPElwert1939,NIMPRSBSeltzer1985}:
\begin{equation}
f_\mathrm{E} = \frac{\beta_1}{\beta_2} \frac{1-\exp\left(-2\pi Z \alpha_f/\beta_1 \right)}{1-\exp\left(-2\pi Z \alpha_f/\beta_2 \right)}\,.
\end{equation}
Figure~\ref{fig:cs_br_nr} displays the Elwert-corrected cross section, $k d\sigma_\mathrm{Br,nr,TFD}/dk$ (red curve), as a function of the normalized photon energy
$k/(\gamma_1-1)$ for a $100\,\mathrm{keV}$ energy electron interacting with neutral Cu atoms (where the Debye shielding vanishes). On the same graph are
plotted the reference tabulated data of Seltzer and Berger \cite{ADNDTSeltzer1986}, $\sigma_\mathrm{SB}$ (black curve), and the nonscreened cross section for
a point Coulomb potential (formula 3BN in Ref.~\onlinecite{RMPKoch1959}), $\sigma_\mathrm{3BN}$ (cyan curve). The difference between $kd\sigma_\mathrm{SB}/dk$
and $k d\sigma_\mathrm{3BN}/dk$ is most pronounced at high and low photon energies owing to, respectively, to Coulomb and screening effects.
By contrast, we observe that the Elwert-corrected screened cross section satisfactorily reproduces Seltzer and Berger's data over the full photon energy range.
For completeness, we have measured the relative error, averaged over photon energies  $0<k/(\gamma_1-1)\le 1$ and in the $1-500\,\rm keV$ kinetic energy range,
between $k d\sigma_\mathrm{Br,nr,TFD}/dk$ and $k d\sigma_\mathrm{SB}/dk$. This error is maximum ($ 78\,\%$) for $1\,\rm keV$ electrons, decreases down to $ 17\,\%$
for $50\,\rm keV$ electrons, and  rises again to $ 53\,\%$  for $500\,\rm keV$ electrons. This difference is expected to increase in higher-$Z$ materials
for which the Born approximation ($2\pi Z/137\beta_1 \ll 1$) becomes less valid.

\begin{figure}[tbh]
\includegraphics[width=0.45\textwidth]{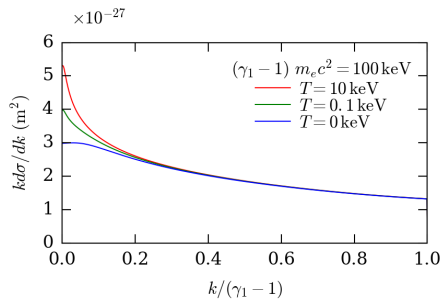}
\caption{Non-relativistic Bremsstrahlung cross section $kd\sigma_\mathrm{Br,nr,TFD}/dk$ vs normalized photon energy $k/(\gamma_1-1)$ for a $100\,\rm keV$ electron in a 
solid-density Cu plasma of temperature ranging from $T=0$ to $10\,\rm keV$. 
\label{fig:cs_br_nr_io}}
\end{figure}

Figure~\ref{fig:cs_br_nr_io} quantifies the effect of the Debye shielding on $k d\sigma_\mathrm{Br,nr,TFD}/dk$ as a function of the target temperature (and corresponding ionization). 
The electron kinetic energy is still set to $100\,\rm keV$, while the temperature of the solid-density Cu target is varied in the range $0 \le T \le 10\,\rm keV$. 
The observed behavior is consistent with the prediction of Fig.~\ref{fig:screening_study} that screening effects weaken at high photon energies:
while, as expected, all curves tend to coincide at high photon energies, the cross section $k d\sigma/dk$ at low photon energies [$k/(\gamma_1-1)\lesssim 0.1$] is
seen to rise (by up to $\sim 60\,\%$ for $k \to 0$) with the target temperature. This variation originates from the increasing effective screening length, which
evolves from $L_\mathrm{TF}$ at low temperatures to $L_\mathrm{D}$ at high temperatures. As $L_\mathrm{D}$ is about one order of magnitude larger than  $L_\mathrm{TF}$
(see Fig.~\ref{fig:screening_study}), a strongly ionized plasma allows for electron-nucleus interactions at larger impact parameters, thus increasing the cross
section. Since screening corrections mainly concern the low-energy side of the spectrum, their impact on the total radiative stopping power remains weak: the latter
increases by a mere $7\%$ increase when the temperature is rised from $0$ to $10\,\rm keV$.

\subsubsection{Moderately relativistic electrons} \label{subsubsec:relativistic}

For moderately relativistic electrons ($2\leq \gamma_1 \leq 100$), the photon-energy-differential Bremsstrahlung cross section is given by the following expression,
valid for arbitrary screening \cite{PCPSBethe1934,RMPKoch1959}:
\begin{align} \label{eq:csre}
\frac{d\sigma_\mathrm{Br,r,TFD}}{dk} & = \frac{4Z^2r_e^2 \alpha_f}{k}
\left\{ \left[ 1+\left(\frac{\gamma_1-k}{\gamma_1} \right)^2\right] \left[ I_1\left(\delta \right) +1 \right] \right. \nonumber \\
& \left.-\frac{2}{3}\frac{\gamma_1-k}{\gamma_1} \left[ I_2 \left(\delta \right) + \frac{5}{6}\right] \right\} \,,
\end{align}
where the functions $I_1$ and $I_2$ account for screening effects:
\begin{align}
I_1(\delta) &= \int_\delta^1 \frac{du}{u^3} \left( u-\delta \right)^2 \left[1-F_e(u)\right]^2 \,, \label{eq:defphi1} \\
I_2(\delta) &= \int_\delta^1 \frac{du}{u^4} \left[ u^3-6\delta^2 u\ln (u/\delta)+3\delta^2 u -4\delta^3 \right] \nonumber \\
& \times \left[ 1-F_e(u) \right]^2 . \label{eq:defphi2}
\end{align}
The argument $\delta=k/2\gamma_1 (\gamma_1-k)$ approximately measures the minimum momentum transfer to the atom in the limit $\gamma_1,\gamma_2\gg 1$.
The above functions involve the atomic form factor (see Appendix~\ref{sec:app_cs_br_re_deriv_anyd})
\begin{equation}
 1- F_e(u) = \frac{2\pi^2}{Z} u^2 \widetilde{V}(u) \,. \label{eq:defF} 
\end{equation}
For a simple single-exponential atomic potential, $V(r)=(q/r)\exp(-r/L)$, the integrals $I_1$ and $I_2$ can be exactly calculated
(see Appendix~\ref{sec:app_cs_br_re_deriv_anyd}):
\begin{align}
I_1 &= \bar{L} \delta \left(\arctan \left(\delta \bar{L}\right) - \arctan \bar{L}\right) - \frac{\bar{L}^2}{2}\frac{\left(1-\delta \right)^2}{1+\bar{L}^2} \nonumber \\
& + \frac{1}{2}\ln \left(\frac{1+\bar{L}^2}{1+\bar{L}^2\delta^2}\right) \, , \label{eq:deriv_I1} \\
2 I_2 & = 4 \bar{L}^{3}\delta^{3}\left(\arctan \bar{L}\delta - \arctan \bar{L}\right) \nonumber \\
&  +\left(1+3 \bar{L}^2 \delta^{2}\right) \ln \left( \frac{1+\bar{L}^2}{1+\bar{L}^2\delta^2} \right)+\frac{6\bar{L}^4\delta^2}{1+\bar{L}^2}\ln\delta \nonumber \\
& +\frac{\bar{L}^2\left(\delta-1\right)\left(\delta+1-4\bar{L}^2\delta^2\right)}{1+\bar{L}^2} \,. \label{eq:deriv_I2}
\end{align}
For the more general atomic potential $V_\mathrm{TFD}(r)$, in contrast to $I_1$, $I_2$ cannot be expressed in terms of elementary functions. 
However, drawing upon Ref.~\onlinecite{RMPTsai1974}, an approximate analytical expression can be derived by matching the asymptotic expressions of
Eq.~\eqref{eq:csre} obtained in the limit $\delta \to 0$ using the double-exponential potential $V_\mathrm{TFD}(r)$ and a single-exponential potential (hereafter
referred to as the ``reduced potential''), $V_\mathrm{R}(r)$. To this goal, we make use of the asymptotic expression
\begin{align}\label{eq:inte}
I &= \lim_{\delta\rightarrow 0}I_ 1 = \lim_{\delta\rightarrow 0}I_ 2 = \int_0^1 u^{3}\left(\frac{1-F_e\left(u\right)}{u^{2}}\right)^{2}\, du \, .
\end{align}
This integral can be analytically evaluated for $V_\mathrm{TFD}$,  as detailed in Appendix~\ref{sec:app_cs_br_re_deriv_seld}.
Let $I_\mathrm{TFD}$ denote its solution:
\begin{align} \label{eq:ITFD}
I_\mathrm{TFD} &= \frac{q_\mathrm{TF}^2}{2q^{2}}\frac{\left(1+\bar{L}_\mathrm{TF}^2 \right) \ln \left(1+\bar{L}_\mathrm{TF}^2 \right)- \bar{L}_{\rm TF}^2}{1+\bar{L}_\mathrm{TF}^2} \nonumber \\
& + \frac{q_\mathrm{D}^2}{2q^{2}} \frac{\left(1+\bar{L}_\mathrm{D}^2 \right) \ln \left( 1+\bar{L}_\mathrm{D}^2 \right)-\bar{L}_{\rm D}^{2}} {1+\bar{L}_\mathrm{D}^2} \nonumber \\
& + \frac{q_\mathrm{TF} q_\mathrm{D}}{q^{2}} \frac{\bar{L}_\mathrm{D}^2 \ln \left(1+\bar{L}_\mathrm{TF}^2 \right)-\bar{L}_\mathrm{TF}^2 \ln \left(1+\bar{L}_\mathrm{D}^2 \right)}{\bar{L}_\mathrm{D}^{2}- \bar{L}_\mathrm{TF}^{2}} \,.
\end{align}
For the reduced potential $V_\mathrm{R}(r) = (q /r) \exp \left(-r/L_\mathrm{R} \right)$, where $L_\mathrm{R}$ is the sought-for reduced screening length, 
the solution to the above integral is
\begin{equation}\label{eq:IRE}
I_\mathrm{R}= \frac{q_{\rm R}^2}{2q^2}\frac{\left(1+\bar{L}_\mathrm{R}^2 \right)\ln\left(1+\bar{L}_\mathrm{R}^2 \right)- \bar{L}_{\rm R}^2}{1+\bar{L}_\mathrm{R}^2} \,. 
\end{equation}
The asymptotic equality $\lim_{\delta \to 0} d\sigma_\mathrm{Br,r,TFD}/dk = \lim_{\delta \to 0} d\sigma_\mathrm{Br,r,R}/dk$ implies $I_\mathrm{TFD} = I_\mathrm{R}$, which defines
the equation solved by $L_\mathrm{R}$. Setting $a=I_\mathrm{TFD}$, $q_{\rm R}=q$ and $x=\bar{L}_\mathrm{R}^2$, this equation can be recast as 
\begin{equation}\label{eq:reduit}
 2a \left( x+1 \right) = \left( x+1 \right) \ln\left( x+1 \right) - x \, .
\end{equation}
The solution involves the Lambert $W$-function:
\begin{equation}\label{eq:lambert-W}
\bar{L}_\mathrm{R} \equiv \sqrt{x} = \left\{ \exp \left[ W \left(-e^{-1-2a}\right)+1+2a\right] - 1 \right\}^{1/2} \, .
\end{equation}
As the coefficient $a$ is positive, $W$ varies over the interval $\left[-1/e,0\right]$, so that $\bar{L}_\mathrm{R}$ is well defined. Combining Eqs.~\eqref{eq:csre}, \eqref{eq:IRE}
and \eqref{eq:lambert-W} gives a closed form analytical expression for the cross section $d\sigma_\mathrm{Br,r,R}/dk$.

\begin{figure}[tbh!]
\centering
\includegraphics[width=0.48\textwidth]{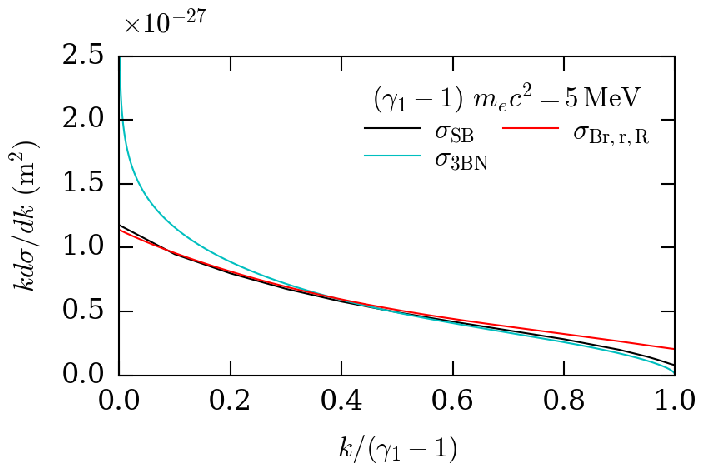}
\caption{Relativistic Bremsstrahlung cross section ($k d\sigma/dk$) vs normalized photon energy for a $5\,\mathrm{MeV}$ electron interacting with neutral Cu atoms. 
Comparison between the reduced analytical formula
($kd\sigma_{\rm Br,r,R}/dk$, red curve), Seltzer and Berger's data \cite{ADNDTSeltzer1986} ($\sigma_\mathrm{SB}$, black curve), and the nonscreened formula 3BN \cite{RMPKoch1959}
($\sigma_\mathrm{3BN}$, cyan curve).}
\label{fig:cs_br_re}
\end{figure}

In Fig.~\ref{fig:cs_br_re}, we compare the Bremsstrahlung cross sections computed using the reduced analytical formula ($k d\sigma_\mathrm{Br,r,R}/dk$, plain red curve) with Seltzer and Berger's reference data \cite{ADNDTSeltzer1986}
($kd\sigma_\mathrm{SB}/dk$, black curve). The case of a $5\,\rm MeV$ electron energy and neutral ($Z^*=0$) Cu atoms is considered. Note that in this limiting case of neutral atoms, the reduced potential ($k d\sigma_\mathrm{Br,r,R}/dk$) and the one numerically computed from the Thomas-Fermi-Debye potential ($k d\sigma_\mathrm{Br,r,TFD}/dk$) exactly coincide.
Importantly, good agreement is found with Ref.~\onlinecite{ADNDTSeltzer1986}, except near $k/(\gamma_1-1)=1$, where a factor of $\sim 2$ discrepancy is observed.
Comparison with the unscreened relativistic cross section ($k d\sigma_{\mathrm{3BN}}/dk$, cyan curve) confirms that shielding by bound electrons is mostly influential at
low photon energies. Overall, the relative error, averaged over $0 < k /\left(\gamma_1-1\right) < 1$, between $\sigma_\mathrm{Br,r,R}/dk$
and $d\sigma_\mathrm{SB}/dk$ is found to steadily drop from $\sim 49\,\%$ for $1\,\rm MeV$ electrons to $\sim 26\,\%$ for $50 \,\rm MeV$ electrons.

\begin{figure}[tbh!]
\centering
\includegraphics[width=0.48\textwidth]{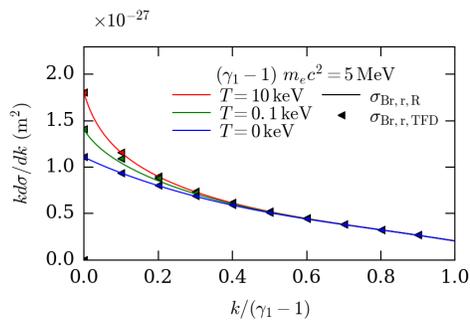}
\caption{Relativistic Bremsstrahlung cross section ($kd\sigma/dk$) vs normalized photon energy for a $5\,\rm MeV$ electron in a solid-density Cu plasma
of temperature ranging from $T=0$ to $10\,\rm keV$. Comparison of the numerically computed cross section using the Thomas-Fermi-Debye potential
($kd\sigma_\mathrm{Br,r,TFD}/dk$, symbols) with the reduced analytical formula ($kd\sigma_\mathrm{Br,r,R}/dk$, solid lines).}
\label{fig:cs_br_re_io}
\end{figure}

Figure~\ref{fig:cs_br_re_io} shows the thermal dependence of the Bremsstrahlung cross section in a solid-density Cu plasma. The incident electron
energy is again set to $5\,\mathrm{MeV}$. For temperature values $0 \le T \le 10\, \rm keV$, one can see that the analytical cross section from the reduced potential
(solid lines, $kd\sigma_{Br,r,R}/dk$) closely reproduces that numerically computed with the Thomas-Fermi-Debye potential (symbols, $kd\sigma_\mathrm{Br,r,TFD}/dk$)
over the full range of photon energies.
As in the nonrelativistic regime (Fig.~\ref{fig:cs_br_nr_io}), Debye screening proves to be mainly significant at low relative photon energies, causing a $\sim 60\,\rm \%$
increase in the cross section when $k \to 0$. Still, this only translates to a quite modest $12\,\%$ enhancement of the total radiated stopping power. Slightly larger
enhancements are predicted with higher-energy electrons.

\subsubsection{Ultra-relativistic electrons} \label{subsubsec:ultra_relativistic}

For ultra-relativistic electron energies ($\gamma_1>100$), the accuracy of the Born-approximation formula \eqref{eq:csre} can be improved by adding the Coulomb correction
term $f_{\mathrm{C}}\left(Z\right)$ as follows:~\cite{RMPKoch1959}:
\begin{align}  \label{eq:csur}
\frac{d\sigma_{Br,ur,TFD}}{dk} & =  \nonumber \\
\frac{4Z^2r_e^2 \alpha_f}{k} &  \left\{ \left[ 1+\left(\frac{\gamma_1-k}{\gamma_1} \right)^2\right] \left[ I_1\left(\delta \right) +1 - f_\mathrm{C}(Z) \right] \right. \nonumber \\
& \left.-\frac{2}{3}\frac{\gamma_1-k}{\gamma_1} \left[ I_2 \left(\delta \right) + \frac{5}{6} - f_\mathrm{C}(Z) \right] \right\} \,,
\end{align}
Introducing the Riemann function $\zeta$, the Coulomb correction term is defined as~\cite{RMPKoch1959}
\begin{align}
f_\mathrm{C} \left(Z\right) &= \frac{\alpha_f^2 Z^2}{1+\alpha_f^2 Z^2} \sum_{n=0}^{\infty} \left(-\alpha_f^2 Z^2 \right)^n \left[ \zeta \left( 2n+1 \right)-1 \right] \, .
\end{align}
In practice, keeping the first four terms has been found sufficient for an accurate computation of $f_\mathrm{C}$ even for high $Z$ values. 
\begin{figure}[tbh!]
\centering
\includegraphics[width=0.48\textwidth]{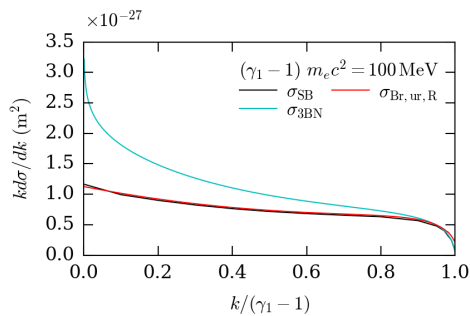}
\caption{Coulomb-corrected ultra-relativistic Bremsstrahlung cross section ($kd\sigma/dk$) vs normalized photon energy for a $100\,\mathrm{MeV}$ electron
interacting with neutral Cu atoms. Comparison between
the reduced analytical formula ($kd\sigma_{\rm Br,ur,R}/dk$, red curve), Seltzer and Berger's data \cite{ADNDTSeltzer1986} ($\sigma_\mathrm{SB}$, black curve) and the
nonscreened formula 3BN \cite{RMPKoch1959} ($\sigma_\mathrm{3BN}$, cyan curve).
\label{fig:cs_br_ur}}
\end{figure}

Figure~\ref{fig:cs_br_ur} shows that, for a $100\,\rm MeV$ energy electron in neutral copper, the Bremsstrahlung cross section based on the reduced potential
($kd\sigma_\mathrm{Br,ur,R}/dk$, plain red curve) agrees well with Berger and Seltzer's data \cite{ADNDTSeltzer1986} ($kd\sigma_\mathrm{SB}/dk$, black curve). Again,
in the present case of neutral atoms, the reduced formula exactly coincides with that evaluated using the full Thomas-Fermi-Debye potential (not shown).
By contrast, the nonscreened ultra-relativistic formula from Ref.~\onlinecite{RMPKoch1959} ($kd\sigma_\mathrm{3BN}/dk$, cyan curve) appears to overestimate the
cross section by a factor of $>1.5$ at photon energies $k/(\gamma_1-1) \lesssim 0.2$. More generally, the relative error, averaged over $0 < k/(\gamma_1-1) < 1$, between
$d\sigma_\mathrm{Br,ur,R}/dk$ and $d\sigma_\mathrm{SB}/dk$) is measured to decrease from $\sim 22\,\%$ for $60 \,\rm MeV$ electrons to $\sim 5 \,\%$ for electron
energies in the range $0.5-10\,\rm GeV$.

\begin{figure}[t]
\centering
\includegraphics[width=0.5\textwidth]{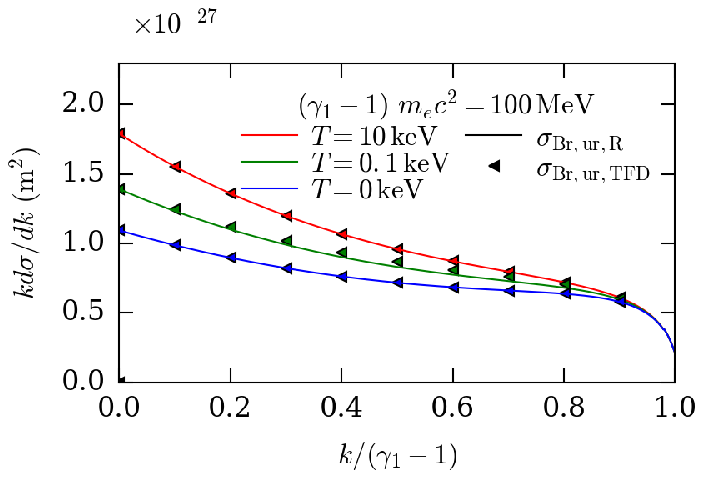}
\caption{Coulomb-corrected, ultra-relativistic Bremsstrahlung cross section ($kd\sigma/dk$) vs normalized photon energy for a $100\,\rm MeV$ electron in a solid-density Cu plasma
of temperature varying from 0 to $10\,\rm keV$. Comparison of the numerically computed cross section using the Thomas-Fermi-Debye potential ($kd\sigma_\mathrm{Br,ur,TFD}/dk$, symbols)
with the reduced analytical formula ($k d\sigma_\mathrm{Br,ur,R}/dk$, solid lines).}
\label{fig:cs_br_ur_io}
\end{figure}

Still for a $100\,\mathrm{MeV}$ energy electron, the thermal variations of the Bremsstrahlung cross section in solid Cu are displayed in Fig.~\ref{fig:cs_br_ur_io} in
the $0-10\,\rm keV$ temperature range. As in the moderately relativistic regime, the analytical formula based on the reduced potential closely matches the numerically evaluated
cross section for all photon energies. In contrast to Fig.~\ref{fig:cs_br_re_io}, though, where Debye screening only affects the emission of relatively low-energy photons
[$k/\left(\gamma_1-1\right) \lesssim 0.2$], it now significantly modifies the cross section up to photon energies $k/\left(\gamma_1-1\right) \lesssim 0.7$.  This increased sensitivity of
the photon spectrum to Debye shielding has a stronger impact on the radiative stoping power, which, in the present case, rises by $\sim 37\,\%$ as the plasma temperature is increased
from 0 to $10\,\rm keV$.

\subsection{Pair production cross sections} \label{subsec:cross_section_pp}


\subsubsection{Bethe-Heitler process} \label{subsubsec:BH}

The cross section of Bethe-Heitler pair production by a photon of normalized energy $k \gg 1$, differential in the normalized positron energy $\gamma_+$,
is given in the ion-rest frame by formula 3D-1003 of Ref.~\onlinecite{RMPMotz1969}:
\begin{align}
\frac{d\sigma_\mathrm{BH,TFD}}{d\gamma_+} & = \frac{4Z^2 r_e^2\alpha_{f}}{k^{3}} \bigg\{ \left( \gamma_+^2 + \gamma_-^2 \right) 
\left[I_1(\delta) + 1 \right] \nonumber \\
& +\frac{2}{3}\gamma_+ \gamma_- \left[ I_2(\delta) + \frac{5}{6} \right] \bigg\} \,, \label{eq:dsdkbh}
\end{align}
where $\delta = k/(2\gamma_+\gamma_-)$. This formula further assumes large electron and positron energies ($\gamma_+,\gamma_- \gg 1$) and negligible nucleus recoil.
We note that it bears much resemblance to the relativistic Bremsstrahlung cross section, Eq.~\eqref{eq:csre}; in particular, it involves the same screening
functions $I_{1,2}$, defined by Eqs.~\eqref{eq:defphi1} and \eqref{eq:defphi2}, which we again evaluate using the Thomas-Fermi-Debye potential,
Eq.~\eqref{eq:V_TFD}, or its reduced form, as defined in Sec.~\ref{subsubsec:relativistic}.

\begin{figure}[t]
\centering
\includegraphics[width=0.49\textwidth]{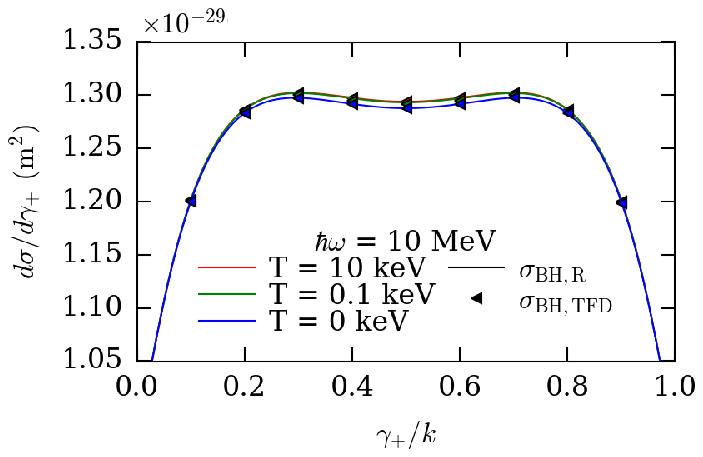}
\includegraphics[width=0.49\textwidth]{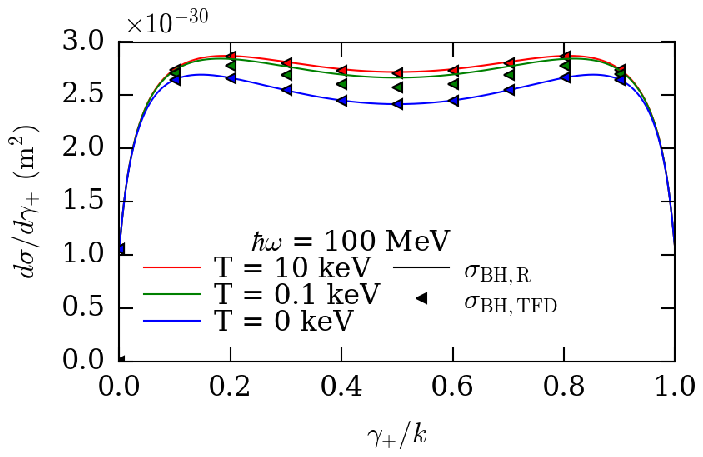}
\caption{Bethe-Heitler cross section vs normalized positron energy $\gamma_+/k$ for a $10\,\mathrm{MeV}$ and $100\,\mathrm{MeV}$ incident photon in a solid-density Cu plasma of
temperature ranging from $T=0$ to $10\,\rm keV$. The cross section computed numerically using the Thomas-Fermi-Debye potential ($d\sigma_\mathrm{BH,r,TFD}/d\gamma_+$, symbols)
closely matches the reduced analytical formula ($d\sigma_\mathrm{BH,r,R}/d\gamma_+$, solid curves). $k$ denotes the normalized photon energy $k=\hbar\omega/mc^2$.}
\label{fig:cs_bh_pp}
\end{figure}

Figure~\ref{fig:cs_bh_pp} illustrates the screening effects on the Bethe-Heitler cross section for a $10$  and $100 \,\rm MeV$ photon incident on a solid-density Cu plasma
in the temperature range $0 \le T \le 10\,\rm keV$.
The cross section has a plateau-like shape symmetric with respect to $\gamma_+ = k/2$, of height (resp. width) decreasing (resp. increasing) with rising photon energy.
For the $100\,\rm MeV$ photon energy considered here, the plateau extends from $\gamma_+ \to 1$  to $\gamma_+ \to k-1$.
Similarly to the Bremsstrahlung, the cross section rises with the plasma temperature/ionization,
and this behavior is more pronounced at large photon energies: the difference between the total cross section at $T=0$ and $10\,\rm keV$
increases from $\sim 0.5\,\%$ for $\hbar\omega \le 10\,\rm MeV$ up to $56\,\%$ for $ \hbar\omega = 10\,\rm GeV$. In the neutral case, we have checked that our formula matches satisfactorily the reference data of Hubbell \textit{et al.}~\cite{JPCRDHubbell1980}. For a photon energy of $10 \, \rm MeV$, the relative difference is $25\%$ while for $100 \, \rm MeV$, it is $8\%$.

\subsection{Trident process} \label{subsubsec:Trident}

The Trident process corresponds to direct pair production by a high-energy electron in the Coulomb field of a nucleus. Due to the lack of tractable analytical formulas, screening
effects on this process will be neglected here. Our approach reproduces that adopted by Vodopiyanov \emph{et al.}~\cite{JGRVodopiyanov2015}. Specifically, we use for the total
Trident cross section the fitting formula provided by Gryaznykh~\cite{JETPLGryaznykh1998}, based on numerical evaluation of the cross section derived by Ba\v{\i}er and
Fadin~\cite{JETPBaier1972}:
\begin{equation}
  \sigma_{\rm T} = 5.22Z^2 \ln^3 \left( \frac{ \gamma_1+4.50 }{6.89} \right) \times 10^{-34} \,\rm m^2 \,,
\end{equation}
where $\gamma_1$ is the incident electron's Lorentz factor. It has recently been suggested \cite{JPPEmbreus2018} that this numerical fit may overestimate
the exact cross section by a factor of $\sim 4$. Yet, in the absence of unambiguous theoretical proof, and in line with Refs.~\onlinecite{PREMyatt2009,JGRVodopiyanov2015},
we continue using it in its original form. The total (normalized) energy of the created pair ($\gamma_p = \gamma_+ +\gamma_-$) is obtained from the singly differential cross
sections calculated by Bhabha~\cite{PRSBhabha1935} in the low- and high-energy limits:
\begin{align}
\frac{d\sigma_{\rm T, nr}}{d\gamma_{p}} & = \frac{\left(Zr_{e}\alpha_{f}\right)^{2}}{32} \times  \left[\log \gamma_{1}^{2} - \frac{161}{60}+C + C_{r} + C_{z}\right] \nonumber \\
& \times \frac{E_{p}^{3}}{\left(m_ec^{2}\right)^{3}}\, , \quad \left(p_{+},p_{-}\right)\ll m_e c  \label{eq:ds_dEpa_part1} \\
\frac{d\sigma_{\rm T, r}}{d\gamma_{p}} & = \frac{56}{9\pi}\left(Zr_{e}\alpha_{f}\right)^{2}\ln\left(\frac{C_{1}E_{p}}{m_ec^{2}}\right) \nonumber \\
& \times \ln\left(\frac{C_{2}m_ec^{2}\gamma_{1}}{E_{p}}\right) \frac{m_ec^2}{E_{p}}  \, , \quad \left(p_{+},p_{-}\right)\gg m_e c  \label{eq:ds_dEpa_part2}
\end{align}
where $C_1$ and $C_2$ are close to unity, and the coefficients $C,C_{r}$ and $C_{z}$ are given by
\begin{align}
& C_{1} = C_{2} = 1 \\
& C = 4\frac{x^{2}}{1-x^{2}}\log \frac{1}{x^{2}}- \frac{4}{3}x^{2}+\frac{1}{6}x^{4} \nonumber \\
& C_{z} = 3 \frac{x^{2}}{1-x^{2}}\left(1-\frac{x^{2}}{1-x^{2}}\log \frac{1}{x^{2}}\right) \nonumber \\
& - \frac{13}{5}x^{2}+ \frac{7}{4}x^{4}-\frac{9}{10}x^{6}+ \frac{1}{5}x^{8} \nonumber \\
& C_{r} = -\frac{3}{2}\frac{x^{2}}{1-x^{2}}\left(1-\frac{x^{2}}{1-x^{2}}\log \frac{1}{x^{2}}\right) \nonumber \\
& + \frac{4}{5}x^{2}- \frac{1}{8}x^{4}-\frac{1}{20}x^{6}+ \frac{1}{40}x^{8} \nonumber
\end{align}
where $x=1/\gamma_{1}$.

An approximate cross section for arbitrary pair energies is obtained by the simple interpolation formula:
\begin{equation}
  \frac{d\sigma_{\rm T}}{d\gamma_p} = \frac{(d\sigma_{\rm T, nr} d\gamma_p)(d\sigma_{\rm T, r} /d\gamma_p)}{d\sigma_{\rm T, nr} /d\gamma_p + d\sigma_{\rm T, r} /d\gamma_p} \,,
\end{equation}
Knowing the pair energy, the positron (or electron) energy is computed making use of the doubly differential cross section (32) of Ref.~\onlinecite{PRSBhabha1935}:
\begin{equation}
  \frac{d\sigma_{\rm T}}{d\gamma_+} \propto \left(\gamma_+^2+\gamma_-^2+\frac{2}{3}\gamma_+\gamma_- \right) \ln \frac{\gamma_+\gamma_-}{\gamma_p} \,,
\end{equation}
where $\gamma_- = \gamma_p - \gamma_+$. As in Ref.~\onlinecite{JGRVodopiyanov2015}, we use this generic shape for the positron/electron distribution in the entire energy
range even though, in principle, this formula holds only for $\gamma_1 \gg (\gamma_+,\gamma_-) \gg 1$.


\section{Monte Carlo implementation} \label{sec:implementation}

\subsection{Pairwise Bremsstrahlung algorithm with arbitrary macro-particle weights} \label{subsec:binary_collision}

We now describe the Monte Carlo implementation of the above photon and pair generation processes in the PIC code \textsc{calder} \cite{NFLefebvre2003, JPCSLobet2016}.
Binary interactions between macro-particles (or macro-photons) are treated using the pairwise collision scheme proposed by Nanbu and Yonemura \cite{JCPNanbu1998}, as has been
done previously for modeling impact ionization \cite{PoPPerez2012}. As the same approach is applied to all processes, we focus in the following on the Bremsstrahlung photon
production.

Let us define two populations of macro-electrons (index $e$) and macro-ions (index $i$). In a given mesh cell,  the (physical) number density of species $\alpha$ is given by
$n_\alpha = \sum_{k=1}^{N_\alpha} W_{\alpha k}$,  where $W_{\alpha k}$ is the numerical weight of the $k$th macro-particle from species $\alpha$, and $N_\alpha$ is the number
of macro-particles of each species in the cell. Bremsstrahlung by electrons colliding with ions in a given cell is modeled by $N_{ei} = \max (N_e, N_i)$ `macro-interactions'
between electron-ion pairs, each pair being randomly sampled with a uniform probability\cite{JCPNanbu1998}. The two macro-particles involved in the $j$th macro-collision 
($1\le j \le N_{ei}$) are characterized by their Lorentz factor ($\gamma_{\alpha j}$), momentum ($\mathbf{p}_{\alpha j}$), velocity ($\boldsymbol{\beta}_{\alpha j}=\mathbf{v}_{\alpha j}/c$),
mass ($m_{\alpha j}$), and the local density of their respective species \cite{JCPNanbu1998, PoPPerez2012}.

Since all the above cross sections are evaluated in the ion rest frame ($K'$), it is convenient to express the electron momentum and energy from the simulation frame $K$ to
$K'$ using the Lorentz transforms~\cite{bookJackson1975}:
\begin{align}
\boldsymbol{p}_e' & =\boldsymbol{p}_e +\left[\frac{\gamma_i -1}{\beta_i^2} \left( \boldsymbol{\beta}_e.\boldsymbol{\beta}_i \right)
- \gamma_i \right] m_e c\gamma_e \boldsymbol{\beta}_e \, , \\
\gamma_e' & = \gamma_e \gamma_i \left( 1 - \boldsymbol{\beta}_e.\boldsymbol{\beta}_i \right) \,,
\end{align}
where the index $j$ is omitted for simplicity. In the following, unprimed (resp. primed) quantities are measured in frame $K$ (resp. $K'$)

Given $\sigma_{\rm B}$ the total Bremsstrahlung cross section for an electron of energy $\gamma_e$, and $v_{\rm rel} = \vert \mathbf{v}_e-\mathbf{v}_i \vert$,
the probability of photon emission during a simulation time-step $\Delta t$ is
\begin{equation}
  \mathcal{P}_{\rm B} = 1 - \exp \left( - n_i \sigma_{\rm B} v_{\rm rel} \Delta t \right) \,. \label{eq:proba1}
\end{equation}  
Making use of the relativistic invariant $\sigma_{\rm B} v_{\rm rel} \gamma_e \gamma_i$~\cite{bookLandau1975}, and noting that $v'_{\rm rel}=v'_e$, the above expression
can be conveniently recast as
\begin{equation} 
  \mathcal{P}_{\rm B} = 1 - \exp\left(-\frac{n_i \sigma'_{\rm B}  \gamma'_e v'_e }{\gamma_e \gamma_i} \Delta t \right) \,. \label{eq:proba2}
\end{equation}
Here $\sigma_{\rm B}'$ denotes the Bremsstrahlung cross section in the ion rest frame, as evaluated in Sec.~\ref{subsec:cross_section_br} in various energy ranges.

Photon emission occurs if $\mathcal{P}_{\rm B} > U$, where $U \in [0;1]$ is a uniform random number. A macro-photon is then created with an energy sampled
from numerical inversion of the cumulative distribution of $d\sigma'_{\rm B}/dk'$. The resulting electron energy is $\gamma'_{ef} = \gamma'_e - k'$. Because $d\sigma'_{\rm B}/dk'$
diverges as $1/k'$ when $k' \to 0$, the $k'$-interval is restricted to $(10^{-n},1)$, where $n \in \mathbb{N}$. The lower bound is chosen small enough that the radiated energy
over the energy interval $(0,10^{-n})$ makes up a negligible fraction (typically $\varepsilon= 10^{-5}$ for $n=7$) of the total radiated energy. The photon is emitted at the electron location,
and parallel to the electron velocity. Finally, we compute the energy and momentum of the photon and decelerated electron in the simulation frame by means of an inverse Lorentz
transformation. 


The radiation probability \eqref{eq:proba2} applies to an electron interacting with an ion population of density $n_i$. Yet in a PIC-Monte Carlo (PIC-MC) simulation,
it has to be modified to take account of the random pairing process of macro-particles with different numerical weights. This is done following the method proposed in
Refs.~\onlinecite{JCPNanbu1998, PoPPerez2012}, whereby one macro-collision represents $\min (W_{ej},W_{ij})$ collisions between real particles. In the event of (physical)
photon emission ($\mathcal{P}_{\rm B} > U$), the macro-photon is created (and the macro-electron is decelerated) with a probability $P_{ej} = W_{ij}/\max(W_{ej},W_{ij})$,
and is then given the macro-electron's statistical weight $W_{ej}$. Introducing $\Delta t_e$, the interaction time-step experienced by real electrons, its cell-averaged value is
\begin{equation}
  \overline{\Delta t}_e = \Delta t_e \frac{ \sum_{j=1}^{N_{ei}} W_{ej} P_{ej}}{n_e}  \equiv \Delta t_e \frac{n_{ei}}{n_e}\, ,
\end{equation}
where we have defined $n_{ei} =\sum_{j=1}^{N_{ei}} \min(W_{ej}, W_{ij})$. Equating $\overline{\Delta t}_e$ with the simulation time-step $\Delta t$ yields $\Delta t_e = n_e \Delta t /n_{ei}$.
The Monte Carlo formulation of Eq.~\eqref{eq:proba2} is then
\begin{equation} 
  \mathcal{P}_{\rm B} = 1- \exp\left(-\frac{n_e n_i}{n_{ei}} \frac{\sigma'_{\rm B}  \gamma'_e v'_e }{\gamma_e \gamma_i} \Delta t \right) \,. \label{eq:proba3}
\end{equation}

A similar scheme is used for the Bethe-Heitler and Trident processes. In the former case, macro-photons and macro-ions are randomly paired in each cell, and the
probability distribution is still of the form \eqref{eq:proba3}, with the difference that $\sigma'_{\rm B}$ is replaced by the (ion rest frame) total Bethe-Heitler cross section
$\sigma'_{\rm BH}$, $n_e$ by the photon density, $n_\gamma$, $\gamma_e$ by $k'\equiv \hbar \omega'/m_ec^2$, and $v'_e$ by $c$. In the event of pair production, the
macro-photon is removed from the simulation, while the energies of the created electron and positron are sampled from the differential cross section $d\sigma'_{\rm BH}/d\gamma'_+$
given in Sec.~\ref{subsubsec:BH}. Both particles are emitted along the photon propagation direction.

\begin{figure}[t]
\centering
\includegraphics[width=0.49\textwidth]{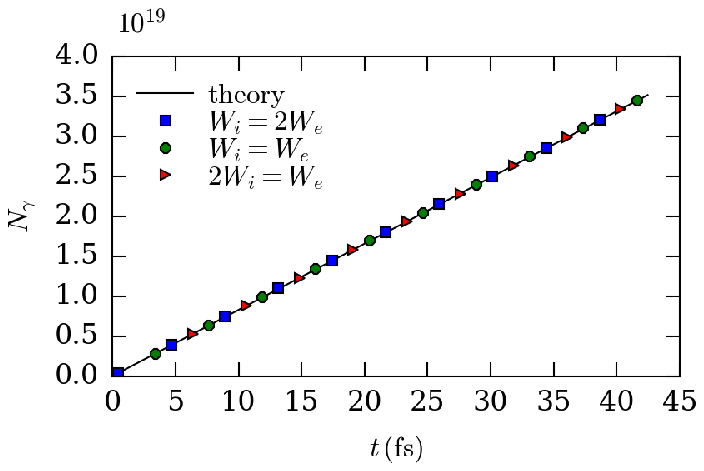}
\includegraphics[width=0.49\textwidth]{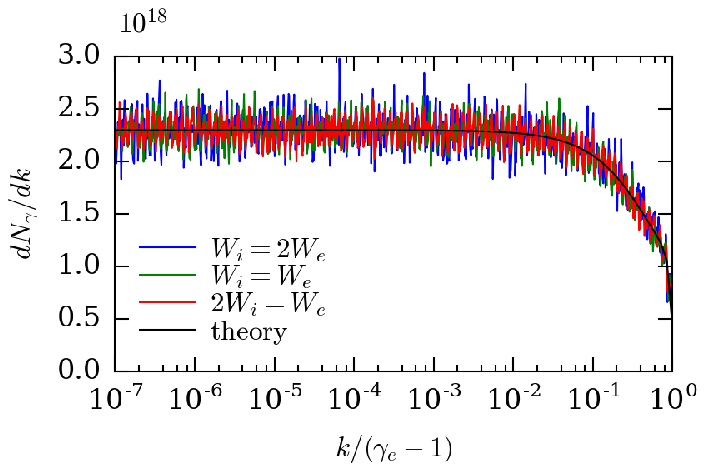}
\caption{(a) Total number and (b) energy spectrum (at $t =42 \, \rm fs$) of Bremsstrahlung photons generated by a $40 \,\rm MeV$ electron beam (Lorentz factor $\gamma_1$)
incident on neutral Cu atoms. Theoretical predictions [Eq.~\eqref{eq:N_br_theo1} in (a) and Eq.~\eqref{eq:N_br_theo1} in (b)] are represented with black lines. Three different
weight ratios are employed: $W_i /W_e =2$ (blue), 1 (green) and $1/2$ (red).
\label{fig:valid_brem_tf}}
\end{figure}

Figures~\ref{fig:valid_brem_tf}(a,b) illustrate the accuracy of our Monte Carlo Bremsstrahlung scheme in the case of a monoenergetic electron beam propagating
through a neutral Cu medium. The beam electrons and background atoms are characterized by $1.6$-$\rm \mu m$-long, uniform density profiles of densities $n_e=10^{21}\,\rm cm^{-3}$
and $n_i=8\times 10^{22}\,\rm cm^{-3}$, respectively. The beam electrons are initialized with an energy of $40\,\rm MeV$ while the ions are taken to be at rest. The simulation,
one-dimensional in configuration space and three-dimensional in velocity space (1D3V), employs periodic boundary conditions. Apart from Bremsstrahlung, all collective and
collisional plasma processes are deactivated. The mesh size is $\Delta x= 160 \, \rm nm$ and the time step is $\Delta t = 0.8 \Delta x$. Each cell contains $100$ macro-electrons,
while the number of macro-ions is varied, from $4000$ to $16000$ per cell, such that the ion-to-electron weight ratio takes on three different values: $W_i/W_e = 2, 1$ and $1/2$.

At early times (when variations in the electron beam energy are negligible), the total number of Bremsstrahlung photons and their energy spectra are expected to vary in time as 
\begin{align}
N_\gamma &= n_e n_i v_e l \sigma_{\rm B} t \,,  \label{eq:N_br_theo1} \\
\frac{dN_\gamma}{dk} &= n_e n_i v_e l  \frac{d\sigma_{\rm B}}{dk} t \,. \label{eq:N_br_theo2}
\end{align}
These evolution laws are well reproduced in Figs.~\ref{fig:valid_brem_tf}(a,b) for the three values of $W_i/W_e$ considered. This validates our Monte Carlo binary interaction
scheme for modeling Bremsstrahlung. Similar validation tests have been performed for the Bethe-Heitler and Trident processes.

\subsection{Collisional-radiative stopping power in a neutral medium} \label{subsec:test}

The \textsc{calder} PIC-MC code now describes all the collisional and radiative processes accounting for the (non-collective) slowing down of fast electrons through a
dense assembly of ions or atoms. To illustrate this increased capability, we have performed a series of simulations of electron beam propagation through a neutral
solid Cu target. These calculations are similar to those presented above except that both
collisional ionization and Bremsstrahlung are taken into account. The electron beam kinetic energy is varied in the range $10^{-2} \le T_1 \le 10^3\,\rm MeV$. For each simulation,
we measure the early-time stopping power (or energy loss rate) resulting from both collisional ionization, $(dT/dx)_{\rm col}$, and Bremsstrahlung, $(dT/dx)_{\rm rad}$. A similar
parametric scan was carried out in Ref.~\onlinecite{PoPPerez2012} in the nonradiative case.

\begin{figure}[h]
\centering
\includegraphics[width=0.5\textwidth]{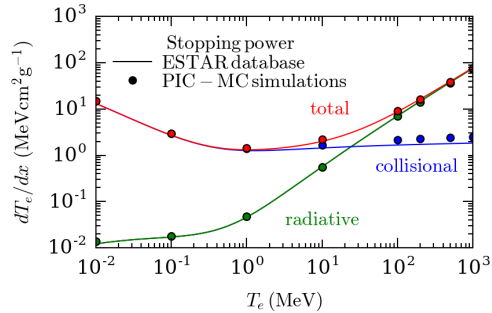}
\caption{Collisional (blue), radiative (green) and total (red) stopping powers of neutral solid copper ($n_i = 6 \times 10^{22}\,\rm cm^{-3}$) as a function of the electron kinetic energy.
Markers show \textsc{calder} PIC-MC simulation results while solid lines plot values from the ESTAR database \cite{ESTAR2017}.}
\label{fig:stpo}
\end{figure}

Figure~\ref{fig:stpo} plots the simulated values of $(dT/dx)_{\rm col}$ (blue markers), $(dT/dx)_{\rm rad}$ (green markers) and of the total stopping power
$(dT/dx)_{\rm tot} = (dT/dx)_{\rm col} + (dT/dx)_{\rm rad}$ (red markers) as a function of $T_1$. Overall, the simulation results closely agree with the reference ESTAR (NIST)
database~\cite{ESTAR2017} (solid lines). For $T_1 \gtrsim 10\,\rm MeV$, however, the PIC-MC simulation tends to overestimate the ionization-induced stopping power, due to
neglect of the density effect \cite{PRSternheimer1966}, whose impact (in reducing the ionization-induced stopping power) increases at large electron energies. This discrepancy
does not affect much the accuracy of the PIC-MC modeled total stopping power because of the prevailing contribution of Bremsstrahlung at energies $\gtrsim 30\,\rm MeV$.

\begin{figure}[t]
\centering
\includegraphics[width=0.5\textwidth]{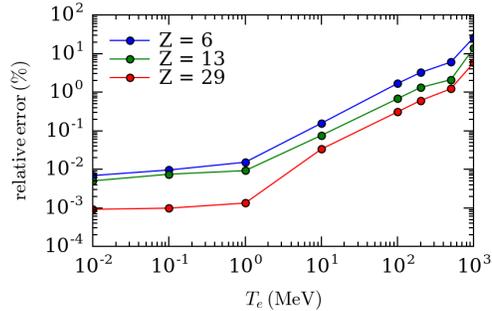}
\caption{Relative error in the PIC-MC collisional stopping power resulting from working in the center-of-mass frame, $\left(dT/dx \right)_{\rm col}^{\rm com}$,
rather than in the ion rest frame, $\left(dT/dx \right)_{\rm col}^{\rm ion}$. as a function of the electron kinetic energy, and for three types of solid materials.
}
\label{fig:frame}
\end{figure}

In the present pairwise algorithm, both Bremsstrahlung and collisional ionization are computed in the ion rest frame, where their respective cross sections are known
and tabulated. By contrast, elastic Coulomb collisions are usually treated in the center-of-mass frame, where they reduce to (identical) rotations of the interacting particles
\cite{PoPSentoku1998}. In Ref.~\onlinecite{PoPPerez2012}, the latter frame was also employed for collisional ionization in order to speed up the calculation. This inconsistency
can be tolerated insofar as the electron kinetic energy is small compared to the ion mass energy; in the opposite limit, it may lead to a significant error. This is shown in
Fig.~\ref{fig:frame}, which plots, for three types of solid materials (C, Al, Cu) and across a range of electron kinetic energies, the relative error in the collisional ionization-induced
stopping power made by working in the center-of-mass frame, $\left(dT/dx \right)_{\rm col}^{\rm com}$ instead of the ion rest frame, $\left(dT/dx \right)_{\rm col}^{\rm ion}$.
For $1\,\rm GeV$ electrons, this error can reach $\sim 20\,\%$ in carbon ($Z=6$) and $\sim 6\,\%$ in copper ($Z=29$). At electron energies $\le 100\,\rm MeV$,
it remains $\lesssim 2\,\%$ for the three atomic elements considered.


\section{Pair generation during relaxation of fast electrons in a finite-size plasma}
\label{sec:generation}

\begin{figure}[t]
\centering
\includegraphics[width=0.49\textwidth]{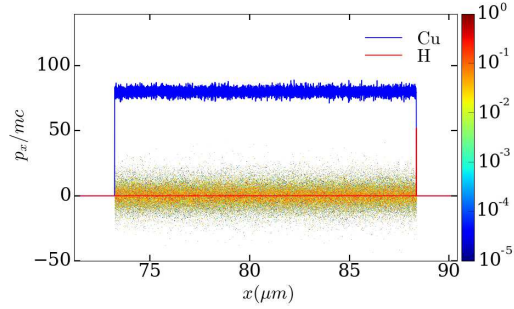}
\caption{1D3V PIC simulation setup: Initial $x-p_x$ phase space of the isotropic fast electron distribution (normalized to its maximum value) in a $15$-$\rm \mu m$-thick
copper target. The blue and red curves plot the density profiles of the bulk Cu$^{5+}$ ions and rear-side contaminant protons.}
\label{fig:setup_calder_myatt_comp}
\end{figure}

We now illustrate the extended capability of the \textsc{calder} PIC code by investigating, in a self-consistent manner, the generation of electron-positron pairs
by fast electrons in micrometer sized solid copper foils. This study is motivated by ongoing experimental research on pair generation by laser-driven
relativistic electrons in solid targets \cite{PRLChen2009, *PRLCHen2010, *PRLChen2015, PoPWilliams2015, *PoPWilliams2016, PRLSarri2013}. Our main goal
is to assess the impact of plasma effects on the competition between the indirect (Bethe-Heitler) and direct (Trident) processes. 



\begin{figure*}[t]
\centering
\includegraphics[width=\textwidth]{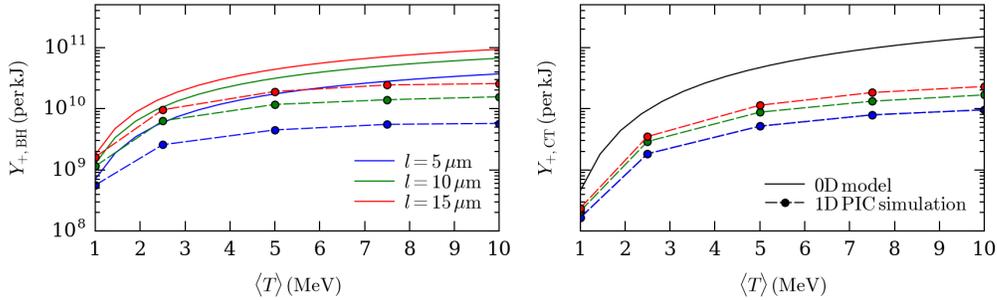}
\caption{(a) Bethe-Heitler and (b) Coulomb Trident positron yield vs average fast-electron kinetic energy for different thicknesses: comparison of the 1D simulation results (dashed curves)
and the predictions of Myatt's 0D theoretical model \cite{PREMyatt2009} (solid curves).}
\label{fig:confront_theo_simu_029}
\end{figure*}

Our 1D3V (1D in configuration space, 3D in velocity space) simulation setup consists of a solid-density Cu plasma slab of thickness varying in the range
$5 \leq l \leq 15 \,\rm \mu m$. The Cu ions are initialized with a uniform density of $8 \times 10^{22}\,\rm cm^{-3}$, a temperature of $1\,\rm eV$,
and an ionization degree $Z^*=5$. The fast electrons have a uniform density profile across the target, and obey a 3D isotropic, relativistic Maxwell-J\"{u}ttner
distribution,
\begin{equation}
  f_e \left(T \right) = \frac{\gamma^2 \beta}{m_e c^2} \frac{\exp\left(-\gamma/\theta \right)}{\theta K_2 \left(1/\theta \right)} \,, \label{eq:mj}
\end{equation}
where $T=(\gamma-1)m_ec^2$, $\theta$ is the temperature normalized to $m_e c^2$, and $K_2$ is a modified Bessel function of the second kind. In the following,
use will be made of the average fast-electron kinetic energy $\langle T \rangle /m_e c^2 =\int f_e (T)\,dT$. For $\theta \ll 1$, $\langle T \rangle/m_ec^2 \simeq 3\theta /2$,
while for $\theta \gg 1$, $\langle T \rangle/m_ec^2 \simeq 3 \theta$. The simulation setup is depicted by the longitudinal $x-p_x$ fast-electron phase space shown in
Fig.~\ref{fig:setup_calder_myatt_comp}.  For $l= 5$, 10 and $15\,\rm \mu m$, the fast-electron density is set to $n_e  = 3 \times 10^{19}\,\rm cm^{-3}$, $1.5 \times 10^{19}\,\rm cm^{-3}$
and $10^{19}\,\rm cm^{-3}$, respectively, which corresponds to a constant areal density. Local charge neutrality is ensured by a bulk electron population of uniform
density $Z^*n_i-n_e$ and $1\,\rm eV$ temperature.  To model the hydrogen-rich surface contaminants, the rear side of the Cu target is coated with a thin
($6.25\,\rm nm$) electron-proton layer of $6\times 10^{22}\,\rm cm^{-3}$ density (red curve). The plasma slab is centered around $x=80\,\rm \mu m$ in a 
$286\,\rm \mu m$-long simulation domain. 

Our choice of a reduced 1D3V geometry is dictated by computational constraints: the spatiotemporal discretization should be fine enough to handle the large
electron density (up to $10^{24}\,\rm cm^{-3}$) of the highly ionized Cu plasma; the integration time should be long enough to reach saturation of the pair generation. 
In order to quench numerical heating, the mesh size is set to $\Delta x = 2.4\,\rm nm$ (so that the simulation box comprises $1.2\times10^5$ cells), and 4th order interpolation
splines are employed. All simulations are run over $2.5\times 10^6$ iterations with time step $\Delta t = 7.4\times 10^{-3}\, \rm fs$. Elastic collisions, impact
ionization, Bremsstrahlung, Bethe-Heitler and Coulomb Trident processes are described. Note that the Bremsstrahlung, Bethe-Heitler and Coulomb Trident processes
are only simulated for the copper ions.

Figures~\ref{fig:confront_theo_simu_029}(a,b) display the total number of positrons generated via the Bethe-Heitler (a) and Coulomb Trident (b) processes as a function of the initial
average fast-electron kinetic energy ($\langle T \rangle = 1-10 \, \mathrm{MeV}$) and the target thickness ($l=5-15 \,\mathrm{\mu m}$). The positron yield is expressed per kJ of
fast-electron kinetic energy. Also plotted are the predictions of the 0D theoretical model proposed by Myatt \emph{et al.} \cite{PREMyatt2009}. This model describes pair generation
from both Bethe-Heitler and Trident processes while taking into account collisional and radiative deceleration of the fast electrons. The finite size of the target only intervenes in its
optical depth for the Bremsstrahlung photons. Perfect confinement of the fast electron in the target is assumed. For the sake of consistency, we have included in this model the
Bremsstrahlung and pair generation cross sections presented in Sec.~\ref{sec:derivation} and considered the Maxwell-J\" uttner energy distribution \eqref{eq:mj} for the fast electrons.

For both pair generation processes, the PIC simulations and 0D model qualitatively agree in predicting an increasing trend in the positron yield with the fast-electron energy
and the target thickness. Besides, they both indicate that the Trident process dominates pair generation at high electron energies and in thin targets. 
PIC simulations predict enhancements of the Bethe-Heitler and Trident yields by factors of $\sim 10$ and $\sim 100$, respectively, when $\langle T \rangle$
is rised from $1\,\rm MeV$ to $10\,\rm MeV$. At $\langle T \rangle = 5\,\rm MeV$ (resp. $10\,\rm MeV$), the Trident process prevails in targets thicker than
$\sim 10\,\rm \mu m$ (resp. $15\,\rm \mu m$). While it gives the correct order of magnitude, the model tends to overestimate the positron yield, especially at large $\langle T \rangle$
and small $l$ for the Bethe-Heitler process. The discrepancy between the model and the PIC results is, however, more pronounced for the Trident process. Also, whereas the
model predicts a Trident positron yield independent of the target thickness, the PIC values increase with $l$, although more slowly than for the Bethe-Heitler process.

The overestimation of the positron generation by the theoretical model stems from the neglect of energy losses associated with ion expansion. As is well-known
\cite{PRLMora2003, *PREMora2005}, fast electrons set up an electrostatic field around the target boundaries. This sheath field, of strength $E_x \propto \sqrt{n_e \langle T \rangle}$,
reflects the electrons back into the target, while accelerating the surface ions outwards. This ion expansion leads to cooling of the fast electrons \cite{PRLMora2003, *PREMora2005}.
This mechanism underlies the generation of energetic ion beams from laser-driven targets of a few $\rm \mu m$ thickness, in which context it is referred to as target normal sheath acceleration (TNSA) \cite{RMPMacchi2013}. 

\begin{figure}[t]
\centering
\includegraphics[width=0.5\textwidth]{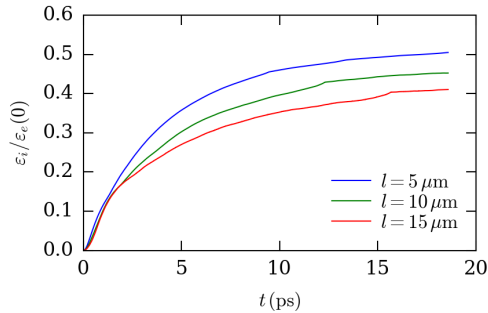}
\caption{Time evolution of the total ion kinetic energy normalized to the initial fast-electron kinetic energy for target thicknesses $l \in(5, 10, 15) \,\rm \mu m$. The electrons
are initially distributed according to a relativistic Maxwell-Juttner law of average kinetic energy $\langle T \rangle = 5\,\rm MeV$.}
\label{fig:eta_ions_5MeV}
\end{figure}

The energy transfer from the fast electrons to the ions is illustrated in Fig.~\ref{fig:eta_ions_5MeV}, which plots the time evolution of the total (copper and hydrogen) ion kinetic
energy normalized to the initial fast-electron energy, $\langle T \rangle = 5 \,\rm MeV$. As expected~\cite{PREMora2005}, the energy conversion efficiency increases
with thinner targets, from $\sim 40\,\%$ at $l=15\,\rm \mu m$ to $\sim 50\,\%$ at $l=15\,\rm \mu m$. Note that such values are likely overestimates owing to the absence of
multidimensional effects (such as transverse dilution of the fast electrons), which should cause earlier saturation of the ion expansion.
 
\begin{figure}[t]
\centering
\includegraphics[width=0.5\textwidth]{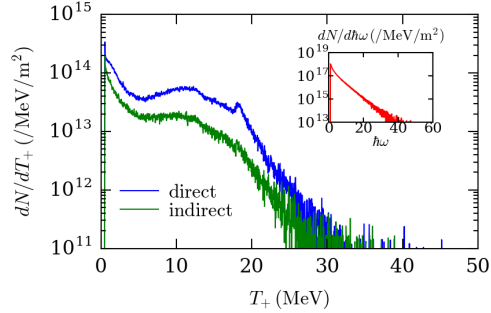}
\caption{Energy spectra of positrons generated via the Bethe-Heitler (green) and Coulomb Trident (blue) processes for $\langle T \rangle = 10\,\rm MeV$ and $l=15\,\rm \mu m$.
The inset shows the corresponding photon energy spectrum.}
\label{fig:spc_illus_L15mum_T100e-1MeV_t520fs}
\end{figure}

In the standard TNSA configuration (\emph{i.e.}, without positron generation), protons usually benefit the most from the fast-electron-induced sheath field due to their largest
charge-to-mass ratio. Here, however, positrons are the lightest positively charged particles, and so respond the fastest to the sheath field while traveling across the target boundaries.
The energy boost that they then undergo can strongly deform their energy spectrum, as evidenced experimentally \cite{PRLChen2009}. The positron energy spectra plotted in
Fig.~\ref{fig:spc_illus_L15mum_T100e-1MeV_t520fs} from both pair generation channels illustrate this mechanism in the case of $\langle T \rangle = 10\,\rm MeV$ and $l = 15\,\rm \mu m$:
both spectra exhibit a bump accompanied by a plateau in the $5-20\,\rm MeV$ range, in contrast to the exponentially decreasing Bremsstrahlung photon spectrum.
\\


\section{Conclusions}
\label{sec:summary}

In summary, we have described a Monte Carlo numerical model for treating Bremsstrahlung, Bethe-Heitler and Coulomb Trident processes in PIC simulations.
For both Bremsstrahlung and Bethe-Heitler processes, we have first derived a set of analytical cross sections suited to arbitrarily ionized media, making use of a
Coulomb atomic potential that reproduces the Thomas-Fermi and Debye-type screened potentials expected, respectively, in the neutral and fully ionized limits.
Overall, those formulas predict an increase in the cross sections at larger plasma temperatures due to a longer-range shielding. We have then developed a binary
collision algorithm adapted to PIC simulations with weighted macro-particles/photons, which we have implemented into the PIC-MC \textsc{calder} code \cite{PoPPerez2012, PRLLobet2015}.
The accuracy of our numerical modeling has been benchmarked against the ESTAR electron stopping power database. Finally,  we have examined the competition
of direct and indirect pair generation channels by a fast-electron population relaxing through a micrometric copper foil of varying thickness. While our 1D simulations qualitatively
support the predictions of the 0D model of Myatt \emph{et al.} \cite{PREMyatt2009}, the latter is shown to overestimate the yields of both pair generation processes
due to the neglect of ion acceleration and of the corresponding electron energy loss. This plasma effect leads to an increasing discrepancy between the PIC and model 
results at larger fast-electron energies and thinner targets. 

Thus upgraded, the PIC-MC \textsc{calder} code is now capable of capturing the full range of radiative (Bremsstrahlung, synchrotron) and pair generation (Bethe-Heitler, Coulomb Trident,
Breit-Wheeler) mechanisms expected to arise in relativistic laser-plasma interactions, notably at next-generation ELI-class laser facilities  \cite{CLEOPapadopoulos2019, OLSung2017, OLZeng2017, MREWeber2017}. 
The numerical investigation of their interplay under various interaction conditions will be the subject of future work.

\section*{Acknowledgments}
The authors acknowledge helpful discussions with O. Embreus and T. F\"ul\"op. PIC simulations were performed using HPC resources at TGCC/CCRT (GENCI project A0010506129).

\appendix 
\section{Non-relativistic Bremsstrahlung cross section with Thomas-Fermi-Debye atomic potential}
\label{sec:app_cs_br_nr_deriv}

We detail here the derivation of the non-relativistic, Bremsstrahlung differential cross-section, Eq.~\eqref{eq:cs_br_nr}. We start by calculating the Fourier transform
\eqref{eq:TF_LDH} of the Thomas-Fermi-Debye potential \eqref{eq:V_TFD}, normalized by $e/\alpha_{f}m_e c^2$
\begin{equation}
  \widetilde{V}_{\rm TFD}\left(u\right) = \frac{Z}{2\pi^2} \left\{\frac{\eta_{\rm TF}\bar{L}_{\rm TF}^2}{1+\bar{L}_{\rm TF}^2 u^2}+\frac{\eta_{\rm D}\bar{L}_{\rm D}^2}{1+\bar{L}_{\rm D}^2 u^2} \right\} \, . \label{eq:app_FT_V_TFD}
\end{equation}
Substituting this expression into Eq.~\eqref{eq:csnr} yields
\begin{align}
& \frac{d\sigma_{\rm NR}}{dk} = \frac{64\pi^4 r_e^2 \alpha_{f}}{3kp_1^2} \left(\frac{Z}{2\pi^2}\right)^2 \times \nonumber \\
& \left\{ \eta_{\rm TF}^2 \bar{L}_{\rm TF}^4\int_{\delta p_{-}}^{\delta p_{+}}\frac{u^3}{\left(1+\bar{L}_{\rm TF}^2 u^2 \right)^2} \, du \right. \nonumber \\
& \left. +\eta_{\rm D}^2 \bar{L}_{\rm D}^4\int_{\delta p_{-}}^{\delta p_{+}}\frac{u^3}{\left(1+\bar{L}_{\rm D}^2 u^2 \right)^2} \, du \right. \nonumber \\
& \left. + 2\eta_{\rm TF}\eta_{\rm D} \bar{L}_{\rm TF}^2\bar{L}_{\rm D}^2\int_{\delta p_{-}}^{\delta p_{+}}\frac{u^3}{\left(1+\bar{L}_{\rm TF}^2 u^2 \right)\left(1+\bar{L}_{\rm D}^2 u^2 \right)} \, du \right\} \, . \label{eq:app_cs_br_nr_02}
\end{align}
Each of these integrals can be analytically solved using standard methods, as detailed in Eqs.~\eqref{eq:appendix_01} and \eqref{eq:appendix_06}. The cross section then
reads
\begin{align}
\frac{d\sigma_{\rm NR}}{dk} & = \frac{16r_e^2 \alpha_{f}Z^2}{3kp_1^2} \times \nonumber \\
& \left\{ \frac{\eta_{\rm TF}^{2}}{2}\left[\ln\left(\frac{1+\bar{L}_{\rm TF}^2 \delta p_{+}^{2}}{1+\bar{L}_{\rm TF}^2\delta p_{-}^{2}}\right)+\frac{1}{1+\bar{L}_{\rm TF}^{2}\delta p_{+}^2}\right. \right. \nonumber \\
& \qquad \qquad \qquad \qquad \left. \left. -\frac{1}{1+\bar{L}_{\rm TF}^2 \delta p_{-}^2}\right] \right.  \nonumber \\
& + \frac{\eta_{\rm D}^{2}}{2}\left[\ln\left(\frac{1+\bar{L}_{\rm D}^2 \delta p_{+}^2}{1+\bar{L}_{\rm D}^2\delta p_{-}^2}\right)+\frac{1}{1+\bar{L}_{\rm D}^{2}\delta p_{+}^{2}}\right. \nonumber \\
& \qquad \qquad \qquad \qquad \left. -\frac{1}{1+\bar{L}_{\rm D}^2 \delta p_{-}^2}\right] \nonumber \\
& \left. + \frac{\eta_{\rm TF}\eta_{\rm D}}{\bar{L}_{\rm TF}^2 -\bar{L}_{\rm D}^2} \left[\bar{L}_{\rm TF}^2 \ln\left(\frac{1+\bar{L}_{\rm D}^2\delta p_{+}^{2}}{1+\bar{L}_{\rm D}^2\delta p_{-}^2}\right)\right. \right. \nonumber \\
& \qquad \qquad \left. \left. -\bar{L}_{\rm D}^2 \ln\left(\frac{1+\bar{L}_{\rm TF}^2 \delta p_{+}^2}{1+\bar{L}_{\rm TF}^2\delta p_{-}^2}\right)\right] \right\} \, , \label{eq:app_cs_br_nr_03}
\end{align}
which can be recast as Eq.~\eqref{eq:cs_br_nr}.

\section{Relativistic Bremsstrahlung cross section with a single-exponential atomic potential} \label{sec:app_cs_br_re_deriv_anyd}

We now consider the derivation of the relativistic Bremsstrahlung cross-section, Eq.~\eqref{eq:csre}. Let us first introduce the electron ($F_e$) and nucleus
($F_n$) form factors:
\begin{equation}
  F_{e,n}(\mathbf{u}) = \frac{1}{Z}\int d^{3}\left(\frac{\mathbf{r}}{r_{\rm C}}\right)\, \frac{r_{\rm C}^{3}\rho_{e,i}(\mathbf{r})}{e}  \exp\left(i \frac{\mathbf{u} \cdot \mathbf{r}}{r_{\rm C}} \right) \,,
\end{equation}
where $\rho_{e,n}(\mathbf{r})$ is the corresponding charge density. Assuming a point-like charge distribution for the nucleus, $\rho_n = Ze \delta (\mathbf{r})$,
gives $F_n = 1$. From Poisson's equation, one can express the atomic form factor as a function of the Fourier transformed of the normalized potential
\begin{equation}
  1 - F_e(u) = \frac{2\pi^2}{Z} u^2 \widetilde{V} (u) \,.\label{eq:app_FT_FF_V}
\end{equation}
Since the atomic potential is taken to be spherically symmetric, the Fourier transformed potential (normalized by $e/4\pi\epsilon_{0}$) can be simplified to
\begin{equation}
  \widetilde{V}(u) = \frac{1}{2\pi^2} \frac{e}{\alpha_f m_e c^2} \frac{1}{u}\int \frac{r}{r_{\rm C}} V(r) \sin (ur/r_{\rm C})  \,  d\left(\frac{r}{r_{\rm C}}\right) .
\end{equation}
In the case of a screened, single-exponential potential, $V(r) = q\exp(-r/L)/r$ with $q=Ze/4\pi\epsilon_0$, one obtains
\begin{equation}
  \widetilde{V}(u) = \frac{Z}{2\pi^2} \frac{\bar{L}^{2}}{1+\bar{L}^{2}u^{2}} \, . \label{eq:app_FT_V}
\end{equation}
where we used the relation $r_e=\alpha_f r_{\rm C}$. We now evaluate the screening correction terms $I_1$ and $I_2$. Using Eqs.~\eqref{eq:app_FT_FF_V}-\eqref{eq:app_FT_V} in Eq.~\eqref{eq:defphi1} yields
\begin{align}
I_1(\delta) & = \bar{L}^4 \int_\delta^1 \frac{u^3}{\left(1+\bar{L}^2 u^2\right)^2} \, du -2 \delta \bar{L}^4 \int_\delta^1 \frac{u^2}{\left(1+\bar{L}^2 u^2 \right)^2} \, du \nonumber \\
& +\delta^2 \bar{L}^4 \int_\delta^1 \frac{u}{\left(1+\bar{L}^2 u^2 \right)^2} \, du \, .
\end{align}
These integrals have closed-form solutions, see Eqs.~\eqref{eq:appendix_01}-\eqref{eq:appendix_03}. There follows
\begin{align}
& I_1(\delta) = \frac{1}{2}\left[\ln \left(\frac{1+\bar{L}^2}{1+\bar{L}^2 \delta^2}\right)+\frac{1}{1+\bar{L}^2}-\frac{1}{1+\bar{L}^2 \delta^2} \right] \\
& - \delta \left[\bar{L}\left[\arctan(\bar{L})-\arctan\left(\bar{L}\delta\right)\right]-\frac{\bar{L}^2}{1+\bar{L}^2}+\frac{\delta \bar{L}^2}{1+\bar{L}^2 \delta^2 }\right] \\
& + \frac{\delta^2 \bar{L}^2}{2} \left[\frac{1}{1+\delta^2 \bar{L}^2}-\frac{1}{1+\bar{L}^2}\right] \,,
\end{align}
which can be readily simplified to Eq.~\eqref{eq:deriv_I1}.

We proceed similarly to calculate $I_2$. Combining Eq.~\eqref{eq:defphi2} with Eqs.~\eqref{eq:app_FT_FF_V}-\eqref{eq:app_FT_V} gives
\begin{align}
 I_2 (\delta) & = \bar{L}^4 \int_\delta^1 \frac{u^3}{\left(1+\bar{L}^2 u^2 \right)^2} \, du \nonumber \\
& - 6 \delta^2 \bar{L}^4 \int_\delta^1 \frac{u\ln u}{\left(1+\bar{L}^2 u^2 \right)^2} \, du \nonumber \\
& + 3\delta^2 \bar{L}^4 \left(2\ln \delta +1\right) \int_\delta^1 \frac{u}{\left(1+\bar{L}^ 2 u^ 2 \right)^2} \, du \nonumber \\
& - 4\delta^3 \bar{L}^4 \int_\delta^1 \frac{1}{\left(1+\bar{L}^2 u^2 \right)^2} \, du \label{eq:defphi2_02} \,.
\end{align}
Again, all of the above integrals can be analytically solved, see Eqs.~\eqref{eq:appendix_01}, \eqref{eq:appendix_03}-\eqref{eq:appendix_05}. One obtains
$I_2 (\delta) = \sum_{i=1}^{4} T_i(\delta) $ where
\begin{align}
T_1(\delta) &= \frac{1}{2}\left[\ln\left(\frac{1+\bar{L}^2}{1+\bar{L}^2\delta^2}\right)+\frac{1}{1+\bar{L}^2}-\frac{1}{1+\bar{L}^2 \delta^2}\right] \,,\nonumber \\ 
T_2(\delta) &= 3\bar{L}^2\delta^2\left[\frac{1}{2}\ln\left(\frac{1+\bar{L}^2}{1+\bar{L}^2\delta^2}\right) + \frac{\bar{L}^2\delta^2 \ln \delta}{1+\bar{L}^2 \delta^2} \right] \,,\nonumber \\ 
T_3(\delta) &= -\frac{3\delta^2 \bar{L}^2}{2}(1+2\ln \delta)\left[\frac{1}{1+\bar{L}^2}-\frac{1}{1+\bar{L}^2\delta^2}\right] \,,\nonumber \\
T_4(\delta) &= 2\delta^{2}\bar{L}^2 \bigg[ \frac{\delta}{1+\bar{L}^2}-\frac{1}{1+\bar{L}^2\delta^2} + 1 - \delta \nonumber \\
& - \delta \bar{L}\left(\arctan\left(\bar{L}\right)-\arctan\left(\bar{L}\delta\right)\right) \bigg]  \,.
\end{align}
Straightforward algebra then leads to the simplified expression Eq.~\eqref{eq:deriv_I2}.

\section{Relativistic Bremsstrahlung cross section with Thomas-Fermi-Debye atomic potential in the $\delta\rightarrow 0$ limit} \label{sec:app_cs_br_re_deriv_seld}

For the two-exponential Thomas-Fermi-Debye potential, Eq.~\eqref{eq:V_TFD}, there is no general analytical expression for $I_2 (\delta)$. However, in the
limit $\delta\rightarrow 0$, both $I_1$ and $I_2$ converge to the same expression $I_{\rm TFD}$  defined by Eq.~\eqref{eq:inte}, which can be exactly calculated. Making
use of Eqs.~\eqref{eq:app_FT_V_TFD} and \eqref{eq:app_FT_FF_V} leads to
\begin{align}
  I_{\rm TFD} & = \frac{q_{\rm TF}^2}{q^{2}} \bar{L}_{\rm TF}^4 \int_0^1  \frac{u^3}{\left(1+\bar{L}_{\rm TF}^2 u^2 \right)^2}\,du \nonumber \\
& + \frac{q_{\rm D}^2}{q^{2}} \bar{L}_{\rm D}^4 \int_0^1 \frac{ u^3}{\left(1+\bar{L}_{\rm D}^2 u^2 \right)^2}\,du \nonumber \\
& + 2 \frac{q_{\rm TF} q_{\rm D}}{q^{2}} \bar{L}_{\rm TF}^2 \bar{L}_{\rm D}^2 \int_0^1 \frac{ u^3}{\left(1+\bar{L}_{\rm TF}^2 u^2 \right)\left(1+\bar{L}_{\rm D}^2 u^2 \right)}\,du \,,
\end{align}
Substituting the closed-form expressions of the above integrals [see Eqs.~\eqref{eq:appendix_01} and \eqref{eq:appendix_06}] directly yields Eq.~\eqref{eq:ITFD}.

\section{Useful indefinite integrals} \label{sec:useful_primitives}

The following indefinite integrals are involved in the above calculations:
\begin{equation}
\int \frac{u^3}{\left(1+\bar{L}^2 u^2 \right)^2}\,du = \frac{1}{2\bar{L}^4}\left[\ln \left(1+\bar{L}^2 u^2 \right)+\frac{1}{1+\bar{L}^2 u^2} \right] \,, \label{eq:appendix_01}
\end{equation}

\begin{equation}
\int \frac{u^2}{\left(1+\bar{L}^2 u^2 \right)^2}\,du = -\frac{1}{2\bar{L}^3}\left[ \frac{\bar{L}u}{1+\bar{L}^2 u^2}-\arctan\left(\bar{L}u \right)\right] \,, \label{eq:appendix_02}
\end{equation}

\begin{equation}
\int \frac{u}{\left(1+\bar{L}^2 u^{2}\right)^2}\,du = -\frac{1}{2\bar{L}^2}\frac{1}{1+\bar{L}^2 u^2} \,, \label{eq:appendix_03}
\end{equation}

\begin{align}
\int \frac{1}{\left(1+\bar{L}^2 u^2 \right)^2}\,du = -\frac{1}{2\bar{L}^2} & \Big[\frac{1}{u\left(1+\bar{L}^2 u^2 \right)}-\frac{1}{u} \nonumber \\
& -\bar{L}\arctan\left(\bar{L}u\right) \Big] \,, \label{eq:appendix_04}
\end{align}

\begin{align}
\int \frac{ u\ln u}{\left(1+\bar{L}^2 u^2 \right)^2}\,du = -\frac{1}{2\bar{L}^2} & \Big[\frac{\ln u}{1+\bar{L}^2 u^2}-\ln u \nonumber \\
& + \frac{1}{2}\ln\left(1+\bar{L}^2 u^2 \right)\Big] \,, \label{eq:appendix_05}
\end{align}
.
\begin{align}
& \int \frac{ u^3}{\left(1+\bar{L}_\alpha^2 u^2 \right)\left(1+\bar{L}_\beta^2 u^2 \right)}\,du \nonumber \\ 
& = \frac{\bar{L}_\alpha^2 \ln\left(1+\bar{L}_\beta^2 u^2 \right)-\bar{L}_\beta^2 \ln\left(1+\bar{L}_\alpha^2 u^2 \right)}{2 \bar{L}_\alpha^2 \bar{L}_\beta^2\left(\bar{L}_\alpha^2 -\bar{L}_\beta^2 \right)} \,, \label{eq:appendix_06}
\end{align}

\bibliographystyle{unsrt}
\bibliography{myBiblio}

\end{document}